\documentclass[11pt,a4paper]{article}

\usepackage{jheppub}
\usepackage[normalem]{ulem}
\usepackage{slashed}

\def\beq{\begin{equation}}
\def\eeq{\end{equation}}
\def\bea{\begin{eqnarray}}
\def\eea{\end{eqnarray}}
\def\nn{\nonumber}
\def\nl{\nonumber\\}

\def\roughly#1{\mathrel{\raise.3ex\hbox
{$#1$\kern-.75em\lower1ex\hbox{$\sim$}}}}
\def\lsim{\roughly<}

\def\s{\sqrt{2}}

\def\bsmumu{b \to s \mu^+ \mu^-}

\def\bctaunu{b \to c \tau^- {\bar\nu}}
\def\bctaunutau{b \to c \tau^- {\bar\nu}_\tau}
\def\bcmunu{b \to c \mu^- {\bar\nu}}
\def\bcmunumu{b \to c \mu^- {\bar\nu}_\mu}
\def\bcenu{b \to c e^- {\bar\nu}}
\def\bcenue{b \to c e^- {\bar\nu}_e}

\def\BDstartaunu{{\bar B}^0 \to D^{*+} \tau^{-} {\bar\nu}_\tau}
\def\BDstarmunu{{\bar B}^0 \to D^{*+} \mu^{-} {\bar\nu}}
\def\BDstarmunumu{{\bar B}^0 \to D^{*+} \mu^{-} {\bar\nu}_\mu}
\def\BDstarlnu{{\bar B}^0 \to D^{*+} \ell^{-} {\bar\nu}_\ell}
\def \cB{{\cal B}}
\def \SM{{\rm SM}}
\def \NP{{\rm NP}}

\def \cM {{\cal M}}
\def \cA {{\cal A}}
\def \ep{\epsilon}
\def \s{\sqrt{2}}
\def \de{\delta}
\def \al{\alpha}
\def \be{\beta}
\def \ga{\gamma}
\def \si{\sigma}
\def \cL{{\cal L}}
\def \cH{{\cal H}}
\def \lp{\left|}
\def \rp{\right|}
\def \lb{\left\{}
\def \rb{\right\}}
\def \ld{\left.}
\def \rd{\right.}
\def \({\left(}
\def \){\right)}
\def \[{\left[}
\def \]{\right]}
\def \Tr{\rm Tr}
\def \ld{\left.}
\def \rd{\right.}
\def \Re{{\rm Re}}

\def \mAp{\mathcal{A}_+}
\def \mAm{\mathcal{A}_-}
\def \mAn{\mathcal{A}_{0}}
\def \mAt{\mathcal{A}_{t}}
\def \mAperp{\mathcal{A}_\perp}
\def \mApar{\mathcal{A}_\parallel}
\def \mAperpT{\mathcal{A}_{\perp, T}}
\def \mAparT{\mathcal{A}_{\parallel, T}}
\def \mA0T{\mathcal{A}_{0, T}}
\def \mDs{m_{D^*}}

\begin{document}

\title{\boldmath CP Violation in ${\bar B}^0 \to D^{*+} \mu^- {\bar\nu}_\mu$}

\author[a]{Bhubanjyoti Bhattacharya,}
\author[b]{Alakabha Datta,}
\author[b]{Saeed Kamali}
\author[c]{and David London}
\affiliation[a]{Department of Natural Sciences, Lawrence Technological University, Southfield, MI 48075, USA}
\affiliation[b]{Department of Physics and Astronomy, \\
108 Lewis Hall, University of Mississippi, Oxford, MS 38677-1848, USA, }
\affiliation[c]{Physique des Particules, Universit\'e de Montr\'eal, \\
C.P. 6128, succ.\ centre-ville, Montr\'eal, QC, Canada H3C 3J7}
\emailAdd{bbhattach@ltu.edu}
\emailAdd{datta@phy.olemiss.edu}
\emailAdd{skamali@go.olemiss.edu}
\emailAdd{london@lps.umontreal.ca}

\abstract{In order to explain the observed anomalies in the
  measurements of $R_{D^{(*)}}$ and $R_{J/\psi}$, a variety of
  new-physics (NP) models that contribute to $\bctaunu$ have been
  proposed. In this paper, we show how CP-violating observables can be
  used to distinguish these NP models. Because ${\vec p}_\tau$ cannot
  be measured (the decay products of the $\tau$ include the undetected
  $\nu_\tau$), obtaining the angular distribution of $\BDstartaunu$ is
  problematic. Instead, we focus here on ${\bar B}^0 \to D^{*+} (\to
  D^0 \pi^+) \mu^- {\bar\nu}_\mu$. This process may also receive
  contributions from the same NP, and LHCb intends to measure the
  CP-violating angular asymmetries in this decay. There are two
  classes of NP models that contribute to $\bcmunumu$. These involve
  (i) a $W'$ (two types) or (ii) a leptoquark (LQ) (six types). The
  most popular NP models predict no CP-violating effects, so the
  measurement of nonzero CP-violating asymmetries would rule them
  out. Furthermore these measurements allow one to distinguish the
  $W'$ and LQ models, and to differentiate among several LQ models.}


\arxivnumber{1903.02567}

\preprint{
{\flushright
UdeM-GPP-TH-19-269 \\
MITP/19-016 \\
}}

\maketitle

\section{Introduction}

At present, there are discrepancies with the predictions of the
standard model (SM) in the measurements of $R_{D^{(*)}} \equiv
\cB(\bar{B} \to D^{(*)} \tau^{-} {\bar\nu}_\tau)/\cB(\bar{B} \to
D^{(*)} \ell^{-} {\bar\nu}_\ell)$ ($\ell = e,\mu$) \cite{RD_BaBar,
  RD_Belle, RD_LHCb, Abdesselam:2016xqt} and $R_{J/\psi} \equiv
\cB(B_c^+ \to J/\psi\tau^+\nu_\tau) / \cB(B_c^+ \to
J/\psi\mu^+\nu_\mu)$ \cite{Aaij:2017tyk}. The experimental results are
shown in Table \ref{tab:obs_meas}. The deviation from the SM in $R_D$
and $R_{D^*}$ (combined) is at the 4$\sigma$ level
\cite{Abdesselam:2017kjf, Bernlochner:2017jka, Bigi:2017jbd,Jaiswal:2017rve},
while it is 1.7$\sigma$ in $R_{J/\psi}$ \cite{Watanabe:2017mip}. These
measurements suggest the presence of new physics (NP) in $\bctaunu$ decays.

\begin{table*}[h]
\begin{center}
\begin{tabular}{|c|c|}
 \hline\hline
Observable & Measurement/Constraint \\
\hline
$R_{D^*}^{\tau/\ell}/(R_{D^*}^{\tau/\ell})_\SM$ & $1.18 \pm 0.06$ \cite{RD_BaBar, RD_Belle, RD_LHCb,Abdesselam:2016xqt} \\
$R_{D}^{\tau/\ell}/(R_{D}^{\tau/\ell})_\SM$ & $1.36 \pm 0.15$ \cite{RD_BaBar, RD_Belle, RD_LHCb,Abdesselam:2016xqt} \\
$R_{D^*}^{\mu/e}/(R_{D^*}^{\mu/e})_\SM$ & $1.00 \pm 0.05$ \cite{Abdesselam:2017kjf} \\
$R_{J/\psi}^{\tau/\mu}/(R_{J/\psi}^{\tau/\mu})_\SM$ & $2.51 \pm 0.97$ \cite{Aaij:2017tyk} \\
 \hline\hline
\end{tabular}
\end{center}
\caption{Measured values of observables that suggest NP in $\bctaunu$.}
\label{tab:obs_meas}
\end{table*}

There have been numerous papers examining the nature of the NP
required to explain the above anomalies. These include both
model-independent \cite{ Watanabe:2017mip, Fajfer:2012jt,
  Datta:2012qk, Tanaka:2012nw, Biancofiore:2013ki, Duraisamy:2013kcw,
  Freytsis:2015qca, Bardhan:2016uhr, Bhattacharya:2016zcw,
  Dutta:2017wpq, Alok:2017qsi, Huang:2018nnq} and model-dependent
analyses \cite{Crivellin:2012ye, Celis:2012dk, He:2012zp, Ko:2012sv,
  Dorsner:2013tla, Sakaki:2013bfa, Greljo:2015mma, Crivellin:2015hha,
  Dumont:2016xpj, Boucenna:2016wpr, Boucenna:2016qad,
  Bhattacharya:2016mcc, Alonso:2016oyd, Celis:2016azn, Wei:2017ago,
  Altmannshofer:2017poe, Buttazzo:2017ixm, Iguro:2017ysu, He:2017bft,
  Biswas:2018jun, Asadi:2018wea, Greljo:2018ogz, Azatov:2018knx,
  Martinez:2018ynq, Fraser:2018aqj, Kumar:2018kmr,
  Robinson:2018gza, Marzo:2019ldg}. There are therefore many possibilities for the
NP. In Refs.~\cite{Sakaki:2012ft, Datta:2012qk, Duraisamy:2014sna,
  Sakaki:2014sea, Bhattacharya:2015ida, Alonso:2016gym, Alok:2016qyh,
  Ligeti:2016npd, Ivanov:2017mrj, Aloni:2017eny, Colangelo:2018cnj,
  Alok:2018uft, Aloni:2018ipm, Asadi:2018sym, Blanke:2018yud, Iguro:2018vqb}, a
variety of observables are proposed for distinguishing the various NP
explanations. These include the $q^2$ distribution, $D^*$
polarization, the $\tau$ polarization, etc. In this paper, we focus on
the measurement of CP-violating observables as a means of
differentiating the NP scenarios\footnote{There are also anomalies in
  various observables involving the decay $\bsmumu$, and several
  different NP explanations have been proposed. In
  Ref.~\cite{Alok:2017jgr} it is shown that these NP models can be
  distinguished through the measurement of CP-violating observables in
  $B \to K^* \mu^+ \mu^-$.}.

All CP-violating effects require the interference of two amplitudes
with different weak (CP-odd) phases. The most common observable is the
direct CP asymmetry, $A_{dir}$, which is proportional to $\Gamma({\bar
  B}^0 \to D^{*+} \tau^- {\bar\nu}_\tau) - \Gamma(B^0 \to D^{*-} \tau^+
\nu_\tau)$. $A_{dir}$ can be nonzero only if the interfering
amplitudes also have different strong (CP-even) phases. Now, strong
phases can only arise in hadronic transitions, and here the only such
transition is ${\bar B} \to D^*$ (or $b \to c$ at the quark
level). Thus, whether the decay proceeds within the SM or with NP, the
strong phase will be the same.  There is one possible exception: If
the NP mediator has colour (e.g., a leptoquark), it can be involved in
gluon exchange, leading to additional strong phases. However, strong
phases generated in this way cannot be large \cite{Datta:2004re}. As a
result, though $A_{dir}$ can be nonzero, we expect it to be small.

The main CP-violating effects in ${\bar B}^0 \to D^{*+} (\to D^0
\pi^+) \tau^- {\bar\nu}_\tau$ therefore appear as CP-violating
asymmetries in the angular distribution\footnote{Another possibility is
to use excited charm mesons, see Ref.~\cite{Aloni:2018ipm}}. These are kinematical
observables, meaning that, in order to generate such effects, the two
interfering amplitudes must have different Lorentz structures.  This
fact allows us to distinguish different NP explanations.

To see this, we note that, in the SM, $\bctaunutau$ arises through the
exchange of a $W$; the four-fermion effective operator is $(V-A)
\times (V-A)$ (LL): $c_\SM \, {\bar c}_L \gamma_\mu b_L {\bar \tau}_L
\gamma^\mu \nu_{\tau,L}$.  If the NP coupling is also LL, it simply
adds to the SM contribution, so that the full coefficient of the
operator is $c_\SM + c_\NP$. Compared to the SM alone, the correction
to the rate is then $2 \, {\rm Re}(c_\SM c^*_\NP) + |c_\NP|^2$. On the
other hand, if the NP four-fermion effective operator has a Lorentz
structure other than LL, there is no SM-NP interference and the
correction to the rate is simply $|c_\NP|^2$. We generally expect NP
effects to be small, i.e., $|c_\NP| < |c_\SM|$, in which case the
largest correction to the rate comes from the SM-NP interference term,
$2 \, {\rm Re}(c_\SM c^*_\NP)$. For this reason, scenarios in which
the NP four-fermion effective operator is also LL are the preferred
explanations. However, in this case, because the SM and NP have the
same Lorentz structure, their interference cannot produce CP-violating
angular asymmetries.  That is, if a nonzero asymmetry were measured,
it would {\it rule out} NP scenarios with purely LL
couplings. Four-fermion effective operators with other Lorentz
structures would be required, and these could be distinguished by the
different types of CP-violating angular asymmetries that they produce.

In Refs.~\cite{Duraisamy:2013kcw, Duraisamy:2014sna}, the decay ${\bar
  B}^0 \to D^{*+} (\to D^0 \pi^+) \tau^- {\bar\nu}_\tau$ was analyzed
in the context of an effective Lagrangian containing NP four-fermion
operators with all Lorentz structures. The angular distribution was
computed, giving the various contributions to the CP-violating angular
asymmetries.  However, there is a practical problem here: the
reconstruction of the angular asymmetries requires the knowledge of
${\vec p}_\tau$. But since the $\tau$ decays to final-state particles
that include $\nu_\tau$, which is undetected, ${\vec p}_\tau$ cannot
be measured.

A complete analysis of CP-violating angular asymmetries in this decay
must therefore include information related to the decay products of
the $\tau$. One such attempt was made in Ref.~\cite{Hagiwara:2014tsa}.
There the decay ${\bar B} \to D \tau^- {\bar\nu}_\tau$ was considered,
with $\tau \to V^- (\to \pi^- \pi^0, \pi^- \pi^+ \pi^- ~{\rm or}~\pi^-
\pi^0\pi^0) \, \nu_\tau$, and a complicated kinematical CP asymmetry
was constructed. Our ultimate goal is to perform a complete angular
analysis of ${\bar B}^0 \to D^{*+} (\to D^0 \pi^+) \tau^-
{\bar\nu}_\tau$, including the angular information from the $\tau$
decay, and compute the NP contributions to all possible CP-violating
angular asymmetries. Some work along these lines can be found in
Ref.~\cite{Alonso:2017ktd}.

In this paper, we take a first step towards this goal by examining the
NP contribution to the CP-violating angular asymmetries in
$\BDstarmunumu$. There are two reasons for starting here. First, LHCb
has announced \cite{Marangotto:2018pbs} that it will perform a
detailed angular analysis of this decay, with the aim of extracting
the coefficients of the CP-violating angular asymmetries. It is
important to show exactly what the implications of these measurements
are for NP. Second, although the preferred explanation of the
$R_{D^{(*)}}$ and $R_{J/\psi}$ anomalies is NP in $\bctaunu$, this
same NP may well also contribute to $\bcmunu$, leading to deviations
from the SM in $\BDstarmunu$.\footnote{Note that, since
  $R_{D^*}^{\mu/e}/(R_{D^*}^{\mu/e})_\SM = 1.00 \pm 0.05$ (Table
  \ref{tab:obs_meas}), NP that contributes to $\bcmunu$ must also
  equally affect $\bcenu$.}

We begin in Sec.~2 with a derivation of the angular distribution for
${\bar B}\to D^* (\to D\pi)\ell^-{\bar\nu}_\ell$, both in the SM and
with the addition of NP. This angular distribution contains several
CP-violating angular asymmetries. In Sec.~3, we describe the various
NP models that can contribute to $\BDstarmunumu$, and compute their
contributions to the various CP-violating observables. This provides
all the NP implications of the measurement of the CP-violating angular
asymmetries. We conclude in Sec.~4.

\section{Angular Analysis}

In this section we discuss the kinematics of the decay ${\bar B}\to
D^* (\to D\pi)\ell^-{\bar\nu}_\ell$ and define the angular observables
in the process using transversity amplitudes. The total decay
amplitude for this process can be expressed as a sum over several
pairs of effective two-body decays. In the most general case, several
of these are due to NP, while one arises from the SM.  We begin by
examining the SM contribution.

\subsection{Transversity amplitudes: SM}
\label{ampsSM}

Following Ref.~\cite{Altmannshofer:2008dz}, the decay ${\bar B}\to D^*
\ell^-{\bar\nu}_\ell$ is considered to be ${\bar B}\to D^* W^{*-}$,
where the on-shell $D^*$ decays to $D\pi$ and the off-shell $W^{*-}$
decays to $\ell^-{\bar\nu}_\ell$\footnote{The angular distributions for
semileptonic $B$ decays were also presented in \cite{Dey:2015rqa}}. Its
amplitude is given by
\bea
\cM_{(m;n)}(B\to D^*W^*) &=& \ep^{*\mu}_{D^*}(m) M_{\mu\nu} \ep^{*\nu}_
{W^*}(n) \label{eq:helicity} ~,~~
\eea
where $\ep^\mu_{V^*}(m)$ is the polarization of a vector particle
($D^*$ or $W^*$). Here $m, n = \pm 1$, 0 and $t$ represent the
transverse, longitudinal and timelike polarizations, respectively.
(Only the off-shell $W^{*-}$ has a timelike polarization.)

In the $B$-meson rest frame we write the polarizations of the two
vector particles as
\bea
\label{poldefs}
& \ep^\mu_{D^*}(\pm) = (0,1,\pm i,0)/\s ~,~~
\ep^\mu_{D^*}(0) = (k_z,0,0,k_0)/m_{D^*} ~, & \\
& \ep^\mu_{W^*}(\pm) = (0,1,\mp i,0)/\s ~,~~
\ep^\mu_{W^*}(0) = -(q_z,0,0,q_0)/\sqrt{q^2} ~,~~
\ep^\mu_{W^*}(t) = q^\mu/\sqrt{q^2} ~, & \nn
\eea
where $k^\mu = (k_0,0, 0,k_z)$ and $q^\mu = (q_0,0,0,q_z)$ are the
four momenta of the $D^*$ and $W^*$, respectively, both written in the
rest frame of the $B$. The polarization vectors of the off-shell $W^*$
satisfy the following orthonormality and completeness relations:
\bea
\ep^{*\mu}_{W^*}(m){\ep_{W^*}}_\mu(m') &=& g_{mm'} ~,~~ \nn\\
\sum\limits_{m,m'}\ep^{*\mu}_{W^*}(m)\ep^\nu_{W^*}(m')g_{mm'} &=& g^{\mu\nu} ~,~~
\eea
where $g_{mm'} = {\rm diag}(+,-,-,-)$ for $m = t,\pm,0$. For the
on-shell $D^*$, these relations are
\bea
\ep^{*\mu}_{D^*}(m){\ep_{D^*}}_\mu(m') &=& -\de_{mm'} ~,~~ \nn\\
\sum\limits_{m,m'}\ep^{*\mu}_{D^*}(m)\ep^\nu_{D^*}(m')\de_{mm'} &=& -g^{\mu\nu} +
\frac{k^\mu k^\nu}{m^2_{D^*}} ~.
\eea

Since the $B$ meson has spin 0, of the 12 combinations of $D^*$ and
$W^*$ polarizations, only 4 are allowed, producing the following
helicity amplitudes:
\bea
\label{helicities}
\cM_{(+;+)}(B\to D^*W^*) &=& \mAp ~,~~ \nn\\
\cM_{(-;-)}(B\to D^*W^*) &=& \mAm ~,~~ \nn\\
\cM_{(0;0)}(B\to D^*W^*) &=& \mAn ~,~~ \nn\\
\cM_{(0;t)}(B\to D^*W^*) &=& \mAt ~.~~
\eea
One may also go to the transversity basis by writing the amplitudes
involving transverse polarizations as
\bea
\cA_{||,\perp} &=& (\mAp \pm \mAm)/\s ~.
\eea

The full amplitude for the decay process $B\to D^*(\to D\pi)\ell^-{\bar\nu}_\ell$ can now
be expressed as
\bea
&& \cM(B\to D^*(\to D\pi)W^*(\to\ell^-{\bar\nu}_\ell)) \\
&& \hskip5truemm \propto
\sum\limits_{m,m'=\pm,0} \ep^{\si}_{D^*}(m) (p_D)_\si \, g_{mm'} \, \ep^{*\rho}_{D^*}(m') \, M_{\rho\nu}
\sum\limits_{n,n'=t,\pm,0} \ep^{*\nu}_{W^*}(n') \, g_{n'n} \, \ep^{\mu}_{W^*}(n) \,
({\bar u}_{\ell}\ga_\mu P_L v_{{\bar\nu}_\ell})
~. \nn
\eea
Here we have made explicit use of the fact that
$\ep^{\si}_{D^*}(p_{D^*})_\si = \ep^{\si}_{D^*}(p_D + p_\pi)_\si = 0$,
so that $A(D^*\to D\pi) \propto \ep^{\si}_{D^*}(p_D - p_\pi)_\si = 2
\ep^{\si}_{D^*}(p_D)_\si$. In the above amplitude, one can project out
the relevant helicity components to obtain
\bea
\label{tempamp}
&& \cM(B\to D^*(\to D\pi)W^*(\to\ell^-{\bar\nu}_\ell)) \nn\\
&& \hskip1truecm \propto
\sum\limits_{m,m'=\pm,0} \, \sum\limits_{n,n'=t,\pm,0}
\ep^{\si}_{D^*}(m) (p_D)_\si \, g_{mm'} \, \cM_{(m',n')}(B\to D^*W^*) \, g_{n'n} \, \ep^{\mu}_{W^*}(n) \,
({\bar u}_{\ell}\ga_\mu P_L v_{{\bar\nu}_\ell}) \nn\\
&& \hskip1truecm \propto
 -~\sum\limits_{m=\pm,0} \, \sum\limits_{n=t,\pm,0} g_{nn}
\cH_{D^*}(m) \, \cM_{(m,n)}(B\to D^*W^*) \, \cL_{W^*}(n)~,~~
\eea
where
\beq
\cH_{D^*}(m) = \ep_{D^*}(m) \cdot p_D ~~,~~~~
\cL_{W^*}(n) = \ep^{\mu}_{W^*}(n) ({\bar u}_{\ell}\ga_\mu P_L v_{{\bar\nu}_\ell}) ~.~~
\eeq
The notation of Eq.~(\ref{tempamp}) can be simplified by defining a
timelike polarization for the $D^*$: $\cH_{D^*}(t) \equiv
\cH_{D^*}(0)$. In this case, the helicities of Eq.~(\ref{helicities})
become $\cM_{(m;m)}(B\to D^*W^*) = \cA_m$ and
\bea
\label{eq:SMamp}
\cM(B\to D^*(\to D\pi)W^*(\to\ell^-{\bar\nu}_\ell)) &\propto&
-~\sum\limits_{m=t,\pm,0} g_{mm} \, \cA_m \, \cH_{D^*}(m) \, \cL_{W^*}(m) ~.~~
\eea

Written in this form, the differential decay rate can now be
constructed from the helicity amplitudes and the Lorentz-invariant
quantities $\cH_{D^*}$ and $\cL_{W^*}$. The spin-summed square of the
amplitude is
\bea
\lp\cM\rp^2 &\propto& \sum\limits_{m,m'=t,\pm,0}g_{mm}g_{m'm'}\(\cA_m\cA^*_{m'}\) \(\cH_
{D^*}(m)\cH^*_{D^*}(m')\)\sum\limits_{\rm spins}\cL_{W^*}(m)\cL^*_{W^*}(m') ~.~~
\label{MVAsquared}
\eea
The leptonic part of the above squared amplitude is given in
Eq.~(\ref{MVAsquaredlep}).

\subsection{New Physics}

From Eq.~(\ref{eq:SMamp}), we see that, in the SM, the decay amplitude
can be written as the product of a hadronic piece $\cH_{D^*}(m)$, a
leptonic piece $\cL_{W^*}(m)$, and a helicity amplitude $\cA_m$,
summed over all helicities $m$. As we will see, this same structure
holds in the presence of NP. We can consider separately the NP
leptonic and hadronic contributions. We begin with the leptonic piece.

In the SM, we have ${\bar B}\to D^* W^{*-}$, where the $W^{*-}$ decays
to $\ell^-{\bar\nu}_\ell$ via a $(V-A)$ interaction. If NP is present,
there are several possible differences. First, there may also be
scalar and/or tensor interactions. Second, the decay products may
include a ${\bar\nu}$ of a flavour other than $\ell$. Finally, a
right-handed (RH), sterile neutrino may be produced
\cite{Asadi:2018wea, Greljo:2018ogz, Robinson:2018gza}. In what
follows, we assume that neutrinos are left-handed, as in the SM,
though we will discuss how our analysis is affected if a RH neutrino
is involved. Regarding the ${\bar\nu}$ flavour, technically we should
write ${\bar\nu}_i$ and sum over all possibilities for $i$ (since the
${\bar\nu}$ is undetected). However, this makes the notation
cumbersome, and does not change the physics.  For this reason, for
notational simplicity, we continue to write ${\bar\nu}_\ell$, though
the reader should be aware that other ${\bar\nu}$ flavours are
possible. Thus, in the presence of NP, the relevant two-body processes
to consider are ${\bar B}\to D^*N^{*-}(\to\ell^-{\bar\nu}_\ell)$,
where $N = S-P, V-A, T$ represent left-handed scalar, vector and
tensor interactions, respectively. In what follows, we label these
$SP$, $VA$ and $T$.  (The $VA$ contribution includes that of the SM.)

Turning to the hadronic piece, we note that the underlying decay is $b
\to c \ell^- {\bar\nu}$. For each of the leptonic $SP$, $VA$ and $T$
Lorentz structures, we introduce NP contributions to the $b \to c$
transition. The effective Hamiltonian is
\bea
{\cal H}_{eff} &=& \frac{G_F V_{cb}}{\sqrt{2}} \Bigl\{
\left[ (1 + g_L) \, {\bar c} \gamma_\mu (1 - \gamma_5) b + g_R \, {\bar c} \gamma_\mu (1 + \gamma_5) b \right]
{\bar \ell} \gamma^\mu (1 - \gamma_5) \nu_\ell \nn\\
&& \hskip-1truecm
+~\left[ g_S \, {\bar c} b + g_P \, {\bar c} \gamma_5 b \right] {\bar \ell} (1 - \gamma_5) \nu_\ell
+ g_T \, {\bar c} \sigma^{\mu\nu} (1 - \gamma_5) b
{\bar \ell} \sigma_{\mu\nu} (1 - \gamma_5) \nu_\ell + h.c. \Bigr\} ~.
\label{4fermi_NP}
\eea

\subsection{Transversity amplitudes: NP}
\label{ampsNP}

Including all possible contributions (SM + NP), the amplitude for the
process can be expressed as
\bea
\cM^{\SM + \NP} &\propto&
\sum_{m,m'=\pm,0} \epsilon_{D^*}^{\nu}(m) (p_{D})_\nu \, g_{mm'} \, \epsilon_{D^*}^{*\mu}(m')
\, M^{SP}_{\mu} \, ({\bar u}_{\ell}P_L v_{{\bar\nu}_\ell}) \nl
&& + \sum\limits_{m,m'} \ep^{\si}_{D^*}(m) (p_{D})_\si \, g_{mm'} \, \ep^{*\rho}_{D^*}(m')
\, M_{\rho\nu}^{VA}
\sum\limits_{n,n'} \ep^{*\nu}_{VA}(n') \, g_{n'n} \, \ep^{\mu}_{VA}(n)
\, ({\bar u}_{\ell}\ga_\mu P_L v_{{\bar\nu}_\ell}) \nl
&& + \sum\limits_{m,m'} \ep^{\beta}_{D^*}(m) (p_{D})_\beta \, g_{mm'} \, \ep^{*\rho}_{D^*}(m') \, M^T_{\rho,\sigma\alpha} \nl
&& \hskip5truemm \times
\sum\limits_{n,n'} \ep^{*\si}_{T}(n') \, g_{n'n} \, \ep^{\mu}_{T}(n)
\sum\limits_{p,p'} \ep^{*\al}_{T}(p') \, g_{p'p} \, \ep^{\nu}_{T}(p)
\, ({\bar u}_{\ell}\sigma_{\mu\nu} P_L v_{{\bar\nu}_\ell}) ~.~~
\label{eq:totalamp}
\eea
The vector part is identical to the SM with the SM coupling replaced
by possible NP couplings in the hadronic amplitudes.

As in the vector-current case, we can define hadronic amplitudes by
contracting the currents with polarization vectors of the intermediate
states. The scalar, vector, and tensor amplitudes are
\bea
\label{eq:SPVAThelicity}
\cM^{SP}_{(m)}(B\to D^* SP^*) &=& \epsilon_{D^*}^{*\mu}(m) \, M^{SP}_{\mu} ~,~~ \nn\\
\cM^{VA}_{(m;n)}(B\to D^* VA^*) &=& \epsilon_{D^*}^{*\mu}(m) \, M^{VA}_{\mu\nu}
\, \ep^{*\nu}_{VA}(n) ~,~~ \nn\\
\cM^T_{(m;n,p)}(B\to D^*T^*) &=& i\ep^{*\rho}_{D^*}(m) \, M^T_{\rho,\si\al}
\, \ep^{*\si}_{T}(n) \, \ep^{*\al}_{T}(p) ~.~~
\eea
Using the above definitions we can now rewrite the total amplitude of
Eq.~(\ref{eq:totalamp}) as
\bea
\cM^{\SM+\NP} &\propto& -~ \sum_{m=\pm,0} \cH_{D^*}(m) \lb \cM^{SP}_{(m)}
\, \cL_{SP} + \sum\limits_{n=t,\pm,0} g_{nn} \, \cM^{VA}_{(m;n)} \, \cL_{VA}(n) \rd ~~~ \nl
&& \hspace{5truecm}\ld + \sum\limits_{n,p=t,\pm,0} g_{nn} \, g_{pp} \, \cM_{(m;n,
p)}^{T} \, \cL_{T}(n,p)\rb ~,~~
\eea
where the leptonic amplitudes have been defined as
\bea
\cL_{SP} &=& {\bar u}_{\ell}P_L v_{{\bar\nu}_\ell} ~,~~ \nn\\
\cL_{VA}(n) &=& \ep^{\mu}_{VA}(n) \, {\bar u}_{\ell}\ga_\mu P_L v_{{\bar\nu}_\ell} ~,~~ \nn\\
\cL_{T}(n,p) &=& -i\ep^{\mu}_{T}(n) \, \ep^{\nu}_{T}(p) \, ({\bar u}_{\ell}\si_{\mu \nu}P_L v_{{\bar\nu}_\ell}) ~.~~
\eea

Since the decaying $B$ meson is a pseudoscalar, conservation of
angular momentum leads to the relationships $m = 0$ for the scalar
part, $m = n$ for the vector part and $m = n + p$ for the tensor
part. In addition, since the tensor current is antisymmetric under the
interchange of $n$ and $p$, the amplitudes corresponding to $n = p$
automatically vanish. Thus, similar to Eq.~(\ref{helicities}), the
non-zero helicity amplitudes in the full angular distribution are
given by
\bea
& \cM^{SP}_{(0)}(B\to D^* SP^*) = \cA_{SP} ~,~~ & \nl
& \cM^{VA}_{(+;+)}(B\to D^* VA^*) = \cA_+ ~,~~ & \nl
& \cM^{VA}_{(-;-)}(B\to D^* VA^*) = \cA_- ~,~~ & \nl
& \cM^{VA}_{(0;0)}(B\to D^* VA^*) = \cA_0 ~,~~ & \nl
& \cM^{VA}_{(0;t)}(B\to D^* VA^*) = \cA_t ~,~~ & \nl
& \cM^T_{(+;+,0)}(B\to D^*T^*) = \cM^T_{(+;+,t)}(B\to D^*T^*) ~=~ \cA_{+,T} ~,~~ & \nl
& \cM^T_{(0;-,+)}(B\to D^*T^*) = \cM^T_{(0;0,t)}(B\to D^*T^*) ~=~ \cA_{0,T} ~,~~ & \nl
& \cM^T_{(-;0,-)}(B\to D^*T^*) = \cM^T_{(-;-,t)}(B\to D^*T^*) ~=~ \cA_{-,T} ~.~~ &
\label{eq:Tampdef}
\eea

The differential decay rate is proportional to the spin-summed
amplitude squared.  We have
\bea
\lp\cM^{\SM + \NP}\rp^2 &=& \lp\cM_{SP}\rp^2 + \lp\cM_{VA}\rp^2 + \lp\cM_T\rp^2 \nn\\
&& \hskip1truecm +~2\Re\[\cM_{SP}\cM^*_{VA} + \cM_{SP}\cM^*_T + \cM_{VA}\cM^*_T\] ~.
\eea
The individual terms are given by
\begin{enumerate}

\item
\bea
|\cM_{SP}|^2 &\propto& \sum_{m,m'=\pm,0} \cM^{SP}_{(m)} \, \cM^{SP*}_{(m')} \, \cH_{D^*}(m)
\, \cH_{D^*}^{*}(m) \sum_{\rm spins}\cL_{SP} \, \cL_{SP}^{*} ~,~~ \nl
&& =\lp\cA_{SP}\rp^2\lp\cH_{D^*}(0)\rp^2\sum_{\rm spins}\cL_{SP} \, \cL_{SP}^{*} ~.
\eea

\item $\lp\cM_{VA}\rp^2$ is given in Eq.~(\ref{MVAsquared}).

\item
\bea
\lp\cM_T\rp^2 &\propto& \sum\limits_{m,m'=\pm,0} \(\cH_{D^*}(m) \, \cH_{D^*}^*(m')\)
\sum\limits_{n,n',p,p'=t,\pm,0} g_{nn} \, g_{n'n'} \, g_{pp} \, g_{p'p'} ~~~ \nl
&& \hspace{1truecm}\times \(\cM_{(m;n,p)}^T \, \cM_{(m';n',p')}^{T*}\) \sum_{\rm spins}
\cL_{T}(n,p) \, \cL_{T}^*(n',p') ~.
\eea

\item
\bea
\cM_{SP}\cM^*_{VA} &\propto& \sum\limits_{m=\pm,0} \cH_{D^*}(0) \, \cH_{D^*}^*(m)\sum\limits
_{n=t,\pm,0} g_{nn} \, \cM_{(0)}^{SP} \nl
&& \hspace{1truecm}\times ~ \cM_{(m;n)}^{VA*}\sum_{\rm spins}\cL_{SP} \, \cL_{VA}^*(n) ~.
\eea

\item
\bea
\cM_{SP}\cM^*_T &\propto& \sum_{m=\pm,0} \cH_{D^*}(0) \, \cH_{D^*}^*(m)\sum\limits_{n,p=t,\pm,0}
g_{nn} \, g_{pp} \, \cM^{SP}_{(0)} \nl
&& \hspace{1truecm}\times ~\cM^{T*}_{(m;n,p)} \sum_{\rm spins} \cL_{SP} \, \cL_T^*(n,p) ~.
\eea

\item
\bea
\cM_{VA} \, \cM^*_T &\propto& \sum\limits_{m,m'=\pm,0} \cH_{D^*}(m) \, \cH_{D^*}^*(m')
\sum\limits_{n,n',p'=t,\pm,0} g_{nn} \, g_{n'n'} \, g_{p'p'} \, \cM^{VA}_{(m;n)} ~~~ \nl
&&\hspace{1truecm} \times
~\cM^{T*}_{(m';n',p')} \sum_{\rm spins}\cL_{VA}(n) \, \cL_T^*(n',p') ~.
\eea

\end{enumerate}

The leptonic contributions to $\lp\cM^{\SM + \NP}\rp^2$ are given in
the appendix, Sec.~\ref{AppendixA}.  The expressions for the helicity
amplitudes in terms of form factors are given in the appendix,
Sec.~\ref{AppendixB}.

The relationships between amplitudes in the helicity and transversity
bases are
\bea
\cA_{||,T} &=& (\cA_{+,T} + \cA_{-,T})/\s ~,~~ \nn\\
\cA_{\perp, T} &=& (\cA_{+,T} - \cA_{-,T})/\s ~.~~
\eea
(A different choice for the transversity basis is used in
Ref.~\cite{Bobeth:2012vn}. However, one can show that the two bases
are equivalent.)

\subsection{Angular Distribution}
\label{AngDist}

In the previous subsection, we computed the square of the full
amplitude for $\bar{B} \to D^* (\to D \pi) \ell^- \bar{\nu}_{\ell}$.
Using Sec.~\ref{AppendixB} in the appendix, this can be expressed as a
function of the final-state momenta. In this section, we obtain the
angular distribution of the decay.

To this end, we use the formalism of helicity angles defined in the
rest frames of the intermediate particles, as shown in Fig.~1. We have
chosen the $z$-axis to align with the direction of the $D^*$ in the
rest frame of the $B$. With this choice of alignment, the helicity
angles $\theta^*$ and $\pi - \theta_\ell$ respectively measure the
polar angles of the $D$ and the charged lepton in the rest frames of
their parent particles ($D^*$ and $N^*$, respectively), and $\chi$ is
the azimuthal angle between the decay planes of the two intermediate
states. For the CP-conjugate decay, the helicity angles are defined in
the same way. Thus, in comparing the decay and the CP-conjugate decay,
${\bar\theta}^* = \theta^*, {\bar\theta}_\ell = \theta_\ell,$ and
${\bar\chi} = \chi$.

\begin{figure}
\begin{center}
\includegraphics[width=0.7\textwidth]{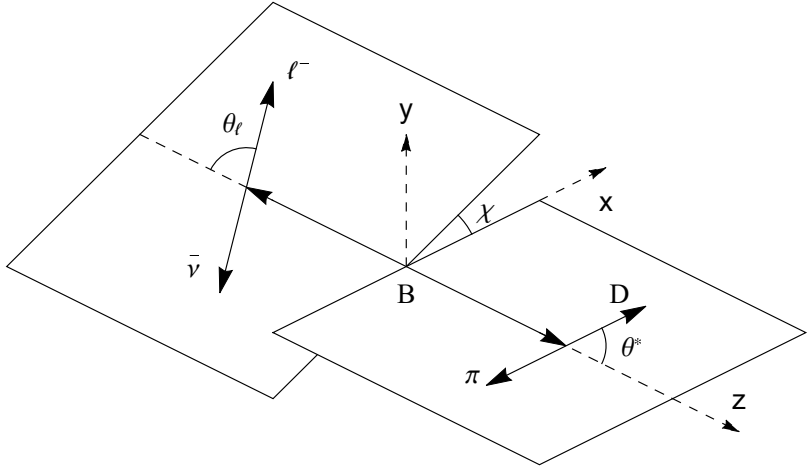}
\caption{Definition of the angles in the $\bar{B} \to D^* (\to D \pi) \ell^- \bar{\nu}_{\ell}$ distribution.}
\end{center}
\label{fig:angdis}
\end{figure}

Using the above definitions we can express the four momenta of the $D$
and the $\ell^-$ in the rest frames of their respective parent
particles as follows:
\bea
p_D^\mu &=& (E_D,\lp\vec p_D\rp\sin \theta^*, 0, \lp\vec p_D\rp\cos\theta^*) ~,~~ \nl
p_\ell^\mu &=& (E_\ell,\lp\vec p_\ell\rp\sin\theta_\ell\cos\chi,\lp\vec p_\ell\rp
\sin\theta_\ell\sin\chi,-\lp\vec p_\ell\rp\cos\theta_\ell) ~,~~
\label{pelldef}
\eea
where $E_X$ and $\vec p_X$ ($X = D, \ell$) represent the energy and
the three-momentum of $X$ in its parent rest frame. The complete
angular distribution can then be written as
\bea
\frac{d^4\Gamma}{dq^2\, d\cos\theta_\ell\, d\cos\theta^*\, d\chi} &=&
 \frac{3}{8 \pi}\frac{G_F^2 |V_{cb}|^2(q^2-m_{\ell}^2)^2 |p_{D^*}|}{2^8 \pi^3 m_B^2 q^2} \nn\\
&& \hskip1truecm
\times ~\mathcal{B}(D^* \to D \pi) \Big(N_1 + \frac{m_\ell}{\sqrt{q^2}} N_2 + \frac{m_\ell^2}{q^2} N_3 \Big) ~,
\eea
where $q = p_\ell + p_{{\bar\nu}_\ell}$, and $|p_{D^*}|=
\sqrt{\lambda(m_B^2, \mDs^2, q^2)}/(2 m_B)$, with $\lambda(a, b,
c)=a^2 + b^2 + c^2 - 2ab - 2ac - 2bc$, is the 3-momentum of $D^*$ in
the $B$-meson rest frame. For $N_1$, $N_2$ and $N_3$, the angular
functions associated with the various (combinations of) helicity
amplitudes are given in Tables \ref{N1}, \ref{N2} and \ref{N3},
respectively. The angular distribution derived here can be compared
with that given in Ref.~\cite{Duraisamy:2014sna}. There are some sign
differences, but these are just conventions -- if everything is
written in terms of form factors, the two angular distributions agree.

\begin{table}[h]
\centering
\begin{tabular}{|l|l|}
\hline
Amplitude in $N_1$                                                                                        & Angular Function \\ \hline
$ |\mAn|^2$                                 &       $4 \sin^2 \theta_\ell \cos^2 \theta^* $              \\ \hline
$|\mAperp|^2$                               &         $ 2 \sin^2 \theta^*(\cos^2 \chi+ \cos^2 \theta_\ell \sin^2\chi)$         \\ \hline
$|\mApar |^2$                               &        $ 2 \sin^2 \theta^*(\cos^2 \theta_\ell \cos^2 \chi+ \sin^2\chi)$             \\ \hline
$ |\mAparT|^2$                              &    $ 32 \sin^2 \theta_\ell \sin^2 \theta^* \cos^2 \chi $                 \\ \hline
$ |\mAperpT|^2$                             &      $32  \sin^2 \theta_\ell \sin^2 \theta^* \sin^2\chi$               \\ \hline
$ |\mA0T|^2$                                &     $ 64 \cos^2 \theta_\ell \cos^2 \theta^*$                \\ \hline
$ |\cA_{SP}|^2$                        &      $4  \cos^2 \theta^*$               \\ \hline
$ {\rm Re}(\mApar \mAperp^*)$                     &     $- 4 \cos\theta_\ell \sin^2 \theta^*$                \\ \hline
$ {\rm Re}(\mAn \mApar^*)$                        &       $- \sqrt{2} \sin 2\theta_\ell \sin 2\theta^* \cos\chi$              \\ \hline
$ {\rm Re}(\mAn \mAperp^*)$                       &    $ 2\sqrt{2} \sin\theta_\ell \sin 2\theta^* \cos\chi$                 \\ \hline
${\rm Re}(\mAparT\cA_{SP}^*)$                &    $8\sqrt{2} \sin\theta_\ell\sin 2\theta^* \cos\chi$                 \\ \hline
${\rm Re}(\mA0T \mAparT^*)$                       &     $16 \sqrt{2}\sin 2\theta_\ell\sin 2\theta^* \cos\chi$                \\ \hline
$ {\rm Re}(\mA0T \cA_{SP}^*)$                &    $32\cos\theta_\ell\cos^2 \theta^*$                 \\ \hline
${\rm Im}(\mAperp \mAn^*)$                        &      $- \sqrt{2} \sin 2\theta_\ell \sin 2\theta^* \sin\chi$               \\ \hline
$ {\rm Im}(\mApar \mAperp^*)$                     &     $ 2 \sin^2 \theta_\ell \sin^2 \theta^* \sin 2\chi$                \\ \hline
$ {\rm Im}(\cA_{SP} \mAperpT^*)$             &       $- 8\sqrt{2} \sin\theta_\ell \sin 2\theta^*\sin\chi$              \\ \hline
$ {\rm Im}(\mAn \mApar^*)$                        &      $- 2\sqrt{2} \sin\theta_\ell \sin 2\theta^* \sin\chi$               \\ \hline
\end{tabular}
\caption{Terms in the $N_1$ part of the angular distribution.}
\label{N1}
\end{table}

\begin{table}[h]
\centering
\begin{tabular}{|l|l|}
\hline
Amplitude in $N_2$                                                                                        & Angular Function \\ \hline
$ {\rm Re}(\mAn \mA0T^*)$                        &     $ -32\cos^2 \theta^*$   \\ \hline
$ {\rm Re}(\mA0T \mAt^*)$                        &     $32\cos\theta_\ell\cos^2 \theta^*$          \\ \hline
$ {\rm Re}(\mAn \cA_{SP}^*)$                &     $-8\cos\theta_\ell\cos^2 \theta^*$            \\ \hline
$ {\rm Re}(\mAt \cA_{SP}^*)$                &     $8\cos^2 \theta^* $  \\ \hline
$ {\rm Re}(\mApar \mAperpT^*) $                  &     $16\cos\theta_\ell\sin^2 \theta^*$  \\ \hline
$ {\rm Re}(\mAparT \mAperp^*)$                   &     $16\cos\theta_\ell\sin^2 \theta^*$   \\ \hline
$ {\rm Re}(\mApar \mAparT^*)$                    &     $-16 \sin^2 \theta^*$  \\ \hline
$ {\rm Re}(\mAperp \mAperpT^*) $                 &     $-16 \sin^2 \theta^*$    \\ \hline
$ {\rm Re}(\mAn \mAperpT^*) $                    &     $-8\sqrt{2}\sin\theta_\ell sin 2\theta^* \cos\chi$ \\ \hline
$ {\rm Re}(\mA0T \mAperp^*)$                     &     $-8\sqrt{2}\sin\theta_\ell\sin 2\theta^* \cos\chi$  \\ \hline
$ {\rm Re}(\mAparT \mAt^*) $                     &     $8\sqrt{2}\sin\theta_\ell\sin 2\theta^* \cos\chi$  \\ \hline
$ {\rm Re}(\mApar \cA_{SP}^*) $             &     $-2\sqrt{2}\sin\theta_\ell\sin 2\theta^* \cos\chi$   \\ \hline
$ {\rm Im}(\mAn \mAparT^* )  $                   &     $8\sqrt{2}\sin\theta_\ell\sin 2\theta^*\sin\chi$     \\ \hline
$ {\rm Im}(\mApar \mA0T^*)$                      &     $-8\sqrt{2}\sin\theta_\ell\sin 2\theta^*\sin\chi$  \\ \hline
$ {\rm Im}(\mAt \mAperpT^*)$                     &     $-8\sqrt{2}\sin\theta_\ell\sin 2\theta^*\sin\chi$  \\ \hline
$ {\rm Im}(\mAperp \cA_{SP}^*) $            &     $-2\sqrt{2}\sin\theta_\ell\sin 2\theta^*\sin\chi$              \\ \hline
\end{tabular}
\caption{Terms in the $N_2$ part of the angular distribution. These
  are suppressed by $m_\ell/\sqrt{q^2}$.}
\label{N2}
\end{table}

\subsection{CP Violation}

The components in the angular distribution that particularly interest
us are those whose coefficients are ${\rm Im}({\mathcal{A}_i}
{\mathcal{A}^*_j})$, where ${\mathcal{A}_{i,j}}$ are two different
helicity amplitudes. These are the terms that are used to generate
CP-violating asymmetries. Note that they are all proportional to
$\sin\chi$.

Technically, these angular components are not, by themselves,
CP-violating observables.  Suppose that the helicity amplitudes
${\mathcal{A}_i}$ and ${\mathcal{A}_j}$ had the same weak phase but
different strong phases.  ${\rm Im}({\mathcal{A}_i}
{\mathcal{A}^*_j})$ would then be nonzero, but this would not indicate
CP violation, since the weak-phase difference vanishes. This would be
a {\it fake} signal. Suppose instead that ${\mathcal{A}_i}$ and
${\mathcal{A}_j}$ had the same strong phase but different weak phases.
${\rm Im}({\mathcal{A}_i} {\mathcal{A}^*_j})$ would again be nonzero,
and in this case it would be a {\it true} CP-violating signal. In
order to distinguish true and fake signals, one must compare the same
quantity in the decay and the CP-conjugate decay. For a true signal,
the angular component will be the same in both decays. This is
because, in going from process to antiprocess, the weak phases change
sign and the azimuthal angle $\chi \to -\chi$. A fake signal will be
indicated if the angular component changes sign. Thus, in the general
case, to obtain a true CP-violating signal, one must add the angular
distributions for the decay and the CP-conjugate decay.
(Even though we are adding the distributions, these are referred to as CP-violating
asymmetries.)  Triple-product asymmetries \cite{TPs, Gronau:2011cf} exhibit a similar
behaviour.  Indeed, the above angular asymmetries are a generalization
of triple products.

\begin{table}[h]
\centering
\begin{tabular}{|l|l|}
\hline
Amplitude in $N_3$                                                                                        & Angular Function \\ \hline
$ |\mAt|^2 $                     &  $ 4\cos^2 \theta^*$     \\ \hline
$ |\mAn |^2$                     &    $4\cos^2 \theta_\ell \cos^2 \theta^*$            \\ \hline
$ |\mAperp |^2$                  &     $ 2 \sin^2 \theta_\ell \sin^2 \theta^* \sin^2\chi$          \\ \hline
$ |\mApar|^2 $                   &   $2\sin^2 \theta_\ell \sin^2 \theta^* \cos^2 \chi $   \\ \hline
$ |\mAparT |^2$                  &   $32\sin^2 \theta^*(\cos^2 \theta_\ell\cos^2 \chi+\sin^2\chi)$    \\ \hline
$ |\mAperpT |^2$                 &    $ 32 \sin^2 \theta^*(\cos^2 \chi+\cos^2 \theta_\ell\sin^2\chi)$  \\ \hline
$ |\mA0T |^2 $                   &      $64\sin^2 \theta_\ell \cos^2 \theta^*$     \\ \hline
$ {\rm Re}(\mAn \mAt^*) $              &     $- 8\cos\theta_\ell \cos^2 \theta^*$        \\ \hline
$ {\rm Re}(\mAn \mApar^*) $            &     $ \sqrt{2}\sin 2\theta_\ell \sin 2\theta^* \cos\chi$  \\ \hline
$ {\rm Re}(\mApar \mAt^*) $            &  $-2\sqrt{2}  \sin\theta_\ell \sin 2\theta^* \cos\chi$   \\ \hline
$ {\rm Re}(\mA0T \mAperpT^*) $         &   $32\sqrt{2}\sin\theta_\ell\sin 2\theta^* \cos\chi$  \\ \hline
$ {\rm Re}(\mA0T \mAparT^*)$           &   $-16\sqrt{2}\sin 2\theta_\ell\sin 2\theta^* \cos\chi$   \\ \hline
$ {\rm Re}(\mAparT \mAperpT^*) $       &      $ -64\cos\theta_\ell\sin^2 \theta^*$   \\ \hline
$ {\rm Im}(\mApar \mAperp^*)$          &    $- 2\sin^2 \theta_\ell \sin^2 \theta^* \sin 2\chi$   \\ \hline
$ {\rm Im}(\mAt \mAperp^*) $           &    $2\sqrt{2} \sin\theta_\ell \sin 2\theta^* \sin\chi$   \\ \hline
$ {\rm Im}(\mAperp \mAn^*)$            &            $\sqrt{2} \sin 2\theta_\ell \sin 2\theta^* \sin\chi$        \\ \hline
\end{tabular}
\caption{Terms in the $N_3$ part of the angular distribution. These
  are suppressed by $m_\ell^2/q^2$.}
\label{N3}
\end{table}

Now, as argued in the introduction, in the case of ${\bar B}\to D^*
(\to D\pi)\ell^-{\bar\nu}_\ell$, the SM and NP contributions all
basically have the same strong phase.  That is, there is no
strong-phase difference between any pair of transversity amplitudes.
In this case, the angular components whose coefficients are ${\rm
  Im}({\mathcal{A}_i} {\mathcal{A}^*_j})$ {\it are} signals of CP
violation.

In Tables \ref{N1}, \ref{N2} and \ref{N3}, one finds, respectively,
four, three and four of these CP-violating observables. However, one
must be careful here. These do not all involve different factors of
${\rm Im}({\mathcal{A}_i} {\mathcal{A}^*_j})$ -- some combinations of
helicity amplitudes appear in more than one Table. Also, these
observables involve only three angular functions, so there can be a
number of different contributions to a single observable. In addition,
the angular components listed in the three Tables are not all the same
size. Compared to Table \ref{N1}, the observables in Tables \ref{N2}
and \ref{N3} are suppressed by $m_\ell/\sqrt{q^2}$ and $m_\ell^2/q^2$,
respectively. Typically, one has $q^2 = O(m_B^2)$, so these
suppression factors are significant. However, if the angular
distribution can be measured in that region of phase space where $q^2
= O(m_\ell^2)$, useful information can be obtained from the
CP-violating observables in these Tables. Finally, the helicity
amplitudes all get contributions from the NP operators in
Eq.~(\ref{4fermi_NP}), so if a particular NP operator is nonzero,
several helicity amplitudes may be affected.

In Table \ref{Tbl:TP} we present all the information about the
CP-violating angular observables: the contributing helicity
amplitudes, the angular functions, the suppression factor, and the NP
couplings probed. This allows us to interpret possible future
measurements.

\begin{table}[]
\centering
\begin{tabular}{|c|c|c|}
\hline
Not suppressed                     & Coupling
& Angular Function \\ \hline
${\rm Im}(\mAperp \mAn^*)$                                   & ${\rm Im}[(1+g_L+g_R)(1+g_L-g_R)^*]$
&      $- \sqrt{2} \sin 2\theta_\ell \sin 2\theta^* \sin\chi$ \\ \hline
${\rm Im}(\mApar \mAperp^*)$                                 & ${\rm Im}[(1+g_L-g_R)(1+g_L+g_R)^*]$
&     $ 2 \sin^2 \theta_\ell \sin^2 \theta^* \sin 2\chi$ \\ \hline
${\rm Im}(\cA_{SP} \mAperpT^*)$                               & ${\rm Im}(g_Pg_T^*)$
&       $- 8\sqrt{2} \sin\theta_\ell \sin 2\theta^*\sin\chi$ \\ \hline
$ {\rm Im}(\mAn \mApar^*)$                                   & ${\rm Im}[(1+g_L-g_R)(1+g_L+g_R)^*]$
&      $- 2\sqrt{2} \sin\theta_\ell \sin 2\theta^* \sin\chi$ \\ \hline \hline
Suppressed by $m_{\ell}/\sqrt{q^2}$ & Coupling
& Angular Function \\ \hline
$ {\rm Im}(\mAn \mAparT^* )  $                                   & ${\rm Im}[(1+g_L-g_R)g_T^*]$
&     $8\sqrt{2}\sin\theta_\ell\sin 2\theta^*\sin\chi$ \\ \hline
$ {\rm Im}(\mApar \mA0T^*)$                                   & ${\rm Im}[(1+g_L-g_R)g_T^*]$
&     $-8\sqrt{2}\sin\theta_\ell\sin 2\theta^*\sin\chi$ \\ \hline
$ {\rm Im}(\mAt \mAperpT^*)$                                 & ${\rm Im}[(1+g_L-g_R)g_T^*]$
&     $-8\sqrt{2}\sin\theta_\ell\sin 2\theta^*\sin\chi$ \\ \hline
${\rm Im}(\mAperp\mathcal{A}_{SP}^*)$                            & ${\rm Im}[(1+g_L+g_R)g_P^*]$
&     $-2\sqrt{2}\sin\theta_\ell\sin 2\theta^*\sin\chi$ \\ \hline \hline
Suppressed by $m_{\ell}^2/q^2$      & Coupling
& Angular Function \\ \hline
${\rm Im}(\mApar \mAperp^*)$                                 & ${\rm Im}[(1+g_L-g_R)(1+g_L+g_R)^*]$
&    $- 2\sin^2 \theta_\ell \sin^2 \theta^* \sin 2\chi$ \\ \hline
$ {\rm Im}(\mAt \mAperp^*) $                                   & ${\rm Im}[(1+g_L+g_R)(1+g_L-g_R)^*]$
&    $2\sqrt{2} \sin\theta_\ell \sin 2\theta^* \sin\chi$ \\ \hline
${\rm Im}(\mAperp\mAn^*)$                                    & ${\rm Im}[(1+g_L+g_R)(1+g_L-g_R)^*]$
&            $\sqrt{2} \sin 2\theta_\ell \sin 2\theta^* \sin\chi$ \\ \hline
\end{tabular}
\caption{The CP-violating terms in the angular distribution, their
  corresponding NP couplings, and the angular functions to which they
  contribute.}
\label{Tbl:TP}
\end{table}

For example, suppose that the angular distribution is measured using
the full data set. In this case, the measurements are dominated by the
unsuppressed contributions of Table \ref{N1}. This angular
distribution contains both CP-conserving and CP-violating pieces, and
both can be affected by NP. We focus on the CP-violating observables
of Table \ref{Tbl:TP}.
\begin{itemize}

\item Suppose that the angular distribution is found to include the
  component $\sin 2\theta_\ell \sin 2\theta^* \sin\chi$.  This
  indicates that ${\rm Im}(\mAperp \mAn^*) \ne 0$, which implies that
  $g_R \ne 0$, and that it has a different (weak) phase than $(1 +
  g_L)$. In this case, one expects to also observe nonzero
  coefficients for the other two angular functions in Table
  \ref{Tbl:TP}, $\sin^2 \theta_\ell \sin^2 \theta^* \sin 2\chi$ and
  $\sin\theta_\ell \sin 2\theta^*\sin\chi$.

\item The third angular function, $\sin\theta_\ell \sin
  2\theta^*\sin\chi$, receives an additional contribution from ${\rm
    Im}(\cA_{SP} \mAperpT^*)$. But if it has been established that
  $g_R \ne 0$, one cannot tell if $g_P$ and $g_T$ are also
  nonzero. This is where the CP-conserving observables come into
  play. From Table \ref{N1}, we see that both $ |\cA_{SP}|^2$ and $
  |\mAperpT|^2$ can be determined from the angular distribution, so in
  principle we will know if they are nonzero (though we will have no
  information about their phases).

\item If it is found that the coefficients of the first two angular
  functions are $\simeq 0$, this implies that $g_R \simeq 0$ (or that
  its phase is the same as that of $(1 + g_L)$). In this case, the
  measurement of a nonzero coefficient of the third angular function
  will point clearly to ${\rm Im}(\cA_{SP} \mAperpT^*) \ne 0$.

\end{itemize}

Finally, suppose that the angular analysis reveals no unsuppressed
CP-violating observables. To probe other such observables, it will now
be necessary to reconstruct the angular distribution for the data with
$q^2 = O(m_\ell^2)$. If this is possible, one can see if the angular
function $\sin\theta_\ell \sin 2\theta^*\sin\chi$ has a nonzero
coefficient in the data suppressed by $m_{\ell}/\sqrt{q^2}$.  If it
does, this indicates that $g_T$ or $g_P$ (or both) is nonzero.  As
noted above, one can perform a cross-check by measuring CP-conserving
observables. In particular, from Table \ref{N1}, we see that the
angular distribution can give us information about new tensor and
scalar interactions.

\section{New-Physics Models}

In Sec.~\ref{AngDist}, we derived the angular distribution for ${\bar
  B}\to D^* (\to D\pi)\ell^-{\bar\nu}_\ell$ in the presence of
NP. This applies to $\ell = e, \mu, \tau$. However, in this paper we
focus specifically on $\BDstarmunumu$, as LHCb intends to perform a
detailed angular analysis of this decay, and measure the CP-violating
observables \cite{Marangotto:2018pbs}. In this section, we examine the
NP models that can generate nonzero CP-violating observables in
$\BDstarmunumu$.

In the SM, the decay $b \to c \ell^- {\bar\nu}$ is due to the
tree-level exchange of a $W$. In order to generate a significant
discrepancy with the SM, the NP contributions to this decay must also
take place at tree level.  There are three classes of NP models in
which this can occur. The NP mediating particle can be a charged Higgs
$H^\pm$ \cite{Crivellin:2012ye, Celis:2012dk, Ko:2012sv,
  Crivellin:2015hha, Celis:2016azn, Wei:2017ago, Biswas:2018jun,
  Altmannshofer:2017poe, Iguro:2017ysu, Azatov:2018knx,
  Martinez:2018ynq, Fraser:2018aqj}, a $W^{\prime \pm}$ boson
\cite{He:2012zp, Greljo:2015mma, Boucenna:2016wpr, Boucenna:2016qad,
  Bhattacharya:2016mcc, Buttazzo:2017ixm, He:2017bft, Asadi:2018wea,
  Greljo:2018ogz, Kumar:2018kmr, Robinson:2018gza}, or a leptoquark
(LQ) \cite{Dorsner:2013tla, Sakaki:2013bfa, Dumont:2016xpj}.

In Ref.~\cite{Alonso:2016oyd}, it was pointed out that there are
important constraints on NP explanations from the $B_c^-$ lifetime. In
particular, NP models with a $H^\pm$ are disfavoured. Below we examine
whether CP-violating observables can be generated in models with a
$W^{\prime \pm}$ or a LQ. Specifically, in each NP model, we determine
which of the NP parameters $g_{L,R,S,P,T}$ [Eq.~(\ref{4fermi_NP})] can
be generated.

We stress that our main goal in this paper is to examine the
implications of the measurement of CP-violating observables in
$\BDstarmunumu$. As such, these $W^{\prime \pm}$ and LQ models are not
complete. That is, there may be constraints from other measurements
that are not taken into account here. For example, it was pointed out
in the introduction that, because
$R_{D^*}^{\mu/e}/(R_{D^*}^{\mu/e})_\SM = 1.00 \pm 0.05$ (Table
\ref{tab:obs_meas}), any NP that contributes to $\bcmunumu$ must
equally affect $\bcenue$. But it is well known that a LQ that couples
to both $\mu$ and $e$ will be constrained by $\mu \to e \gamma$ and $b
\to s e \mu$ \cite{Crivellin:2017dsk}.  Should a CP-violating
observable be measured in $\BDstarmunumu$ suggesting the presence of
LQs, these constraints must be taken into account at the
model-building stage.

\subsection{$W^{\prime\pm}$ Models}

The $W'$ is a vector boson, so it can contribute only to $g_L$ and/or
$g_R$ of Eq.~(\ref{4fermi_NP}). Two classes of $W'$ models have been
proposed. In the first \cite{Greljo:2015mma, Boucenna:2016wpr,
  Boucenna:2016qad, Bhattacharya:2016mcc, Buttazzo:2017ixm,
  Kumar:2018kmr}, the $W'$ is SM-like, coupling only to left-handed
fermions. Thus, this $W'_L$ contributes only to $g_L$, which means
that no CP-violating effects can be generated.

The second class uses LR models: one has a right-handed $W'_R$,
and the decay involves a sterile RH neutrino. The $W'_R$ couples only
to right-handed fermions and so contributes to neither $g_L$ nor $g_R$
(since these operators involve a left-handed neutrino). One can allow
for NP that couples to a RH neutrino by adding the following NP
operators to Eq.~(\ref{4fermi_NP}):
\bea
{\cal H}'_{eff} &=& \frac{G_F V_{cb}}{\sqrt{2}} \Bigl\{
\left[ g'_L \, {\bar c} \gamma_\mu (1 - \gamma_5) b + g'_R \, {\bar c} \gamma_\mu (1 + \gamma_5) b \right]
{\bar \mu} \gamma^\mu (1 + \gamma_5) \nu \nn\\
&& \hskip-1truecm
+~\left[ g'_S \, {\bar c} b + g'_P \, {\bar c} \gamma_5 b \right] {\bar \mu} (1 + \gamma_5) \nu
+ g'_T \, {\bar c} \sigma^{\mu\nu} (1 + \gamma_5) b
{\bar \mu} \sigma_{\mu\nu} (1 + \gamma_5) \nu + h.c. \Bigr\} ~.
\eea
Just as in Table \ref{Tbl:TP}, CP-violating observables can be
produced due to the interference of any two of these NP operators.
However, the $W'_R$ contributes only to $g'_R$, so that, once again,
no CP-violating effects can be generated.

CP-violating observables can be generated if the $W'$ contributes to
both $g_L$ and $g_R$ (or $g'_L$ and $g'_R$ if the neutrino is
RH). This can occur in the LR model if the SM $W$ mixes with the
$W'_R$. However, constraints from $b \to s\gamma$ force this mixing to
be small, $\lsim O(10^{-3})$ \cite{He:2017bft}, which means that any
CP-violating effects are tiny.

Thus, the only way to generate sizeable CP-violating effects is if
there is a $W'_L$ and a $W'_R$, both with large contributions to $b \to
c \ell^- {\bar\nu}$, and there is significant mixing. Such a model has
not yet been proposed, but it is a possibility.

\subsection{Leptoquark Models}

There are ten models in which the LQ couples to SM particles through
dimension $\le 4$ operators \cite{pdg}. These include five spin-0 and
five spin-1 LQs. Six of these can contribute to $\bcmunumu$
\cite{Sakaki:2013bfa}. Three have fermion-number-conserving couplings
and three have fermion-number-violating couplings. The interaction
Lagrangian that generates the contributions to $\bcmunumu$ is given by
\bea
\cL^{\rm LQ} &=& \cL^{\rm LQ}_{F=0} + \cL^{\rm LQ}_{F=-2} ~, \nn\\
\cL^{\rm LQ}_{F=0} &=&
(h^{ij}_{1L}{\bar Q}_{iL}\ga^\mu L_{jL} + h^{ij}_{1R}{\bar d}_{iR}\ga^\mu\ell_{jR})U_{1\mu}
+ h^{ij}_{3L}{\bar Q}_{iL} {\vec \si}\ga^\mu L_{jL} \cdot {\vec U}_{3\mu} \nn\\
&&
+~(h^{ij}_{2L}{\bar u}_{iR} L_{jL} + h^{ij}_{2R}{\bar Q}_{iL} i\si_2\ell_{jR})R_2 + h.c., \nn\\
\cL^{\rm LQ}_{F=-2} &=&
(g^{ij}_{1L}{\bar Q}^c_{iL}i\si_2 L_{jL} + g^{ij}_{1R}{\bar u}^c_{iR}\ell_{jR})S_1
+ (g^{ij}_{3L}{\bar Q}^c_{iL}i\si_2 {\vec \si} L_{jL}) \cdot {\vec S}_3 \nn\\
&&
+~(g^{ij}_{2L}{\bar d}^c_{iR}\ga_\mu L_{jL} + g^{ij}_{2R}{\bar Q}^c_{iL}\ga_\mu\ell_{jR})V^\mu_2 + h.c.
\eea
Here $Q$ and $L$ represent left-handed quark and lepton $SU(2)_L$
doublets, respectively; $u$, $d$ and $\ell$ represent right-handed
up-type quark, down-type quark and charged lepton $SU(2)_L$ singlets,
respectively. The indices $i$ and $j$ are the quark and lepton
generations. $\psi^c = C {\bar\psi}^T$ is a charge-conjugated field.

For all six models, we integrate out the LQ to form four-fermion
operators. We then perform Fierz transformations to put these
operators in the form of Eq.~(\ref{4fermi_NP}). In this way, we
determine which LQs contribute to which $g_{L,R,S,P,T}$
coefficients. \\

\noindent
$U_1$:
\beq
\cL_{\rm LQ}
\supset (h^{22}_{1L}{\bar c}_{L}\ga^\mu \nu_{\mu L} + h^{32}_{1L}{\bar b}_{L}\ga^\mu \mu_{L} + h^{32}_{1R}{\bar
b}_{iR}\ga^\mu\mu_{R})U_{1\mu} + h.c.
\eeq
Four-fermion operators:
\bea
\cL^{\rm eff} &=& -\frac{1}{M^2_{U_1}}\[h^{22}_{1L}h^{32*}_{1L}({\bar c}_{L}\ga^\mu\nu_{\mu L})({\bar\mu}_{L}\ga_\mu
b_{L}) + h^{22}_{1L}h^{32*}_{1R}({\bar c}_{L}\ga^\mu\nu_{\mu L})({\bar\mu}_{R}\ga_\mu
b_{R})\] + h.c.
\eea
Fierz transformation:
\bea
\cL^{\rm eff} &=& -\frac{1}{M^2_{U_1}}\[h^{22}_{1L}h^{32*}_{1L}({\bar c}_{L}\ga^\mu b_L)({\bar\mu}_{L}\ga_\mu
\nu_{\mu L}) - 2 h^{22}_{1L}h^{32*}_{1R}({\bar c}_{L}b_{R})({\bar\mu}_{R}\nu_{\mu L})\] + h.c.
\eea

\noindent
$U_3$:
\bea
\cL_{\rm LQ}
&\supset& (h^{22}_{3L}{\bar c}_{L}\ga^\mu \nu_{\mu L} - h^{32}_{3L}{\bar b}_{L}\ga^\mu \mu_{L})U_{3\mu} + h.c.
\eea
Four-fermion operator:
\bea
\cL^{\rm eff} &=& \frac{1}{M^2_{U_3}} h^{22}_{3L}h^{32*}_{3L}({\bar c}_{L}\ga^\mu\nu_{\mu L})({\bar\mu}_{L}\ga_\mu
b_{L}) + h.c.
\eea
Fierz transformation:
\bea
\cL^{\rm eff} &=& \frac{1}{M^2_{U_3}} h^{22}_{3L}h^{32*}_{3L}({\bar c}_{L}\ga^\mu b_L)({\bar\mu}_{L}\ga_\mu
\nu_{\mu L}) + h.c.
\eea

\noindent
$R_2$:
\bea
\cL_{\rm LQ}
&\supset& (h^{22}_{2L}{\bar c}_{R}\nu_{\mu L} - h^{32}_{2R}{\bar b}_{L}\mu_{R})R_{2} + h.c.
\eea
Four-fermion operator:
\bea
\cL^{\rm eff} &=& \frac{1}{M^2_{R_2}} h^{22}_{2L}h^{32*}_{2R}({\bar c}_{R}\nu_{\mu L})({\bar\mu}_{R}b_{L}) + h.c.
\eea
Fierz transformation:
\bea
\cL^{\rm eff} &=& -\frac{1}{8M^2_{R_2}}\[4h^{22}_{2L}h^{32*}_{2R}({\bar c}_{R}b_L)({\bar\mu}_{R}\nu_{\mu L})
+ h^{22}_{2L}h^{32*}_{2R}({\bar c}_{R}\si^{\mu\nu}b_L)({\bar\mu}_{R}\si_{\mu\nu}\nu_{\mu L})\] + h.c.
\eea

\noindent
$S_1$:
\bea
\cL_{\rm LQ}
&\supset& (g^{22}_{1L}{\bar c}^c_{L}\mu_{L} - g^{32}_{1L}{\bar b}^c_{L}\nu_{\mu L} + g^{22}_{1R}{\bar c}^c_{R}\mu_{R})S_{1} + h.c.
\eea
Four-fermion operators:
\bea
\cL^{\rm eff} &=& \frac{1}{M^2_{S_1}}\[g^{22^*}_{1L}g^{32}_{1L}({\bar b}^c_{L}\nu_{\mu L})({\bar\mu}_{L}c^c_{L}) +
g^{22^*}_{1R}g^{32}_{1L}({\bar b}^c_{L}\nu_{\mu L})({\bar\mu}_{R}c^c_{R})\] + h.c.
\eea
Fierz transformation:
\bea
\cL^{\rm eff} &=& \frac{1}{8M^2_{S_1}}\[4g^{22^*}_{1L}g^{32}_{1L}({\bar c}_{L}\ga^\mu b_{L})({\bar\mu}_{L}\ga_\mu\nu_{\mu L})
- 4 g^{22^*}_{1R}g^{32}_{1L}\rd({\bar c}_Rb_L)({\bar\mu}_{R}\nu_{\mu L}) ~~~ \nl
&&\hspace{2truecm}\ld +~ g^{22^*}_{1R}g^{32}_{1L}({\bar c}_R\si^{\mu\nu}b_L)({\bar\mu}_{R}\si_{\mu\nu}\nu_{\mu L})\] + h.c.
\eea

\noindent
$S_3$:
\bea
\cL_{\rm LQ}
&\supset& -(g^{22}_{3L}{\bar c}^c_{L}\mu_{L} + g^{32}_{3L}{\bar b}^c_{L}\nu_{\mu L})S_{3} + h.c.
\eea
Four-fermion operator:
\bea
\cL^{\rm eff} &=& -\frac{1}{M^2_{S_3}} g^{22^*}_{3L}g^{32}_{3L}({\bar b}^c_{L}\nu_{\mu L})({\bar\mu}_{L}c^c_{L}) + h.c.
\eea
Fierz transformation:
\bea
\cL^{\rm eff} &=& -\frac{1}{2M^2_{S_3}} g^{22^*}_{3L}g^{32}_{3L}({\bar c}_{L}\ga^\mu b_{L})({\bar\mu}_{L}\ga_\mu\nu_{\mu L}) + h.c.
\eea

\noindent
$V_2$:
\bea
\cL_{\rm LQ}
&\supset& (g^{32}_{2L}{\bar b}^c_{R}\ga_\mu\nu_{\mu L} + g^{22}_{2R}{\bar c}^c_{L}\ga_\mu\mu_{R})V^\mu_{2} + h.c.
\eea
Four-fermion operator:
\bea
\cL^{\rm eff} &=& -\frac{1}{M^2_{V_2}} g^{22^*}_{2R}g^{32}_{2L}({\bar b}^c_{R}\ga^\mu\nu_{\mu L})({\bar\mu}_{R}\ga_\mu c^c_{L}) + h.c.
\eea
Fierz transformation:
\bea
\cL^{\rm eff} &=& \frac{2}{M^2_{V_2}} g^{22^*}_{2R}g^{32}_{2L}({\bar c}_{L}b_{R})({\bar\mu}_{R}\nu_{\mu L}) + h.c.
\eea

In Table \ref{4fermi_LQ} we summarize the contributions of all the LQs
to the $g_{L,R,S,P,T}$ coefficients of Eq.~(\ref{4fermi_NP}).

\begin{table}[h]
\begin{center}
\begin{tabular}{|c|c|c|c|c|c|} \hline \hline
Model & $g_L$ & $g_R$ & $g_S$ & $g_P$ & $g_T$ \\ \hline
$U_1$ & $\frac12 {h^{22}_{1L}h^{32*}_{1L}}$ &   0   & $-{h^{22}_{1L}h^{32*}_{1R}}$ & $-{h^{22}_{1L}h^{32*}_{1R}}$ &   0   \\ \hline
$U_3$ & $-\frac12{h^{22}_{3L}h^{32*}_{3L}}$ & 0 & 0 & 0 & 0 \\ \hline
$R_2$ & 0 & 0 & $\frac14{h^{22}_{2L}h^{32*}_{2R}}$ & $-\frac14{h^{22}_{2L}h^{32*}_{2R}}$
& $\frac{1}{16}{h^{22}_{2L}h^{32*}_{2R}}$ \\ \hline
$S_1$ & $-\frac14{g^{32}_{1L}g^{22*}_{1L}}$ & 0 & $\frac14{g^{32}_{1L}g^{22*}_{1R}}$
& $-\frac14 {g^{32}_{1L}g^{22*}_{1R}}$ & $-\frac{1}{16}{g^{32}_{1L}g^{22*}_{1R}}$ \\ \hline
$S_3$ & $\frac14{g^{32}_{3L}g^{22*}_{3L}}$ & 0 & 0 & 0 & 0 \\ \hline
$V_2$ & 0 & 0 & $-{g^{22*}_{2R}g^{32}_{2L}}$ & $-{g^{22*}_{2R}g^{32}_{2L}}$ & 0 \\ \hline \hline
\end{tabular}
\end{center}
\caption{Contributions of the various LQs to the $g_{L,R,S,P,T}$
  coefficients of Eq.~(\ref{4fermi_NP}). All entries must be
  multiplied by $1/(\s G_F V_{cb} M^2_{\rm LQ})$.}
\label{4fermi_LQ}
\end{table}

\subsection{CP Violation}

As shown in Table \ref{Tbl:TP}, the CP-violating observables involve
any pair of $\{ (1 + g_L), g_R, g_P, g_T \}$.  Above we have seen that
the $W'$ and most LQ models contribute to $g_L$. It must be pointed
out that, in $\bcmunumu$, $g_L$ cannot be large. This is because it is
the coefficient of the $(V-A)\times (V-A)$ operator ${\bar c}
\gamma_\mu (1 - \gamma_5) b {\bar \mu} \gamma^\mu (1 - \gamma_5)
\nu_\mu$, which is related by $SU(2)_L \times U(1)_Y$ to the $\bsmumu$
operator ${\bar s} \gamma_\mu (1 - \gamma_5) b {\bar \mu} \gamma^\mu
(1 - \gamma_5) \mu$ \cite{RKRDpaper}. In order to explain the
anomalies in the $\bsmumu$ observables, we require \cite{Alok:2017sui}
\beq
g_L = \frac{\alpha}{2\pi} (-0.68 \pm 0.12) = O(10^{-3}) ~.
\eeq
In $(1 + g_L)$, this is negligible.

Going beyond $g_L$, we note that $g_R$ can only be due to a $W'$, and
$g_P$ and $g_T$ can only be generated in LQ models. Furthermore, not
all $W'$ models lead to a nonzero $g_R$. And not all LQ models produce
$g_P$ and/or $g_T$. Putting all of this together, if NP is present in
$\bcmunumu$, we see that the measurement of CP-violating observables
can give us a great deal of information as to its identity.

First of all, most NP models proposed to explain the $R_{D^{(*)}}$ and
$R_{J/\psi}$ experimental data contribute only to $g_L$ (in
$\bctaunutau$). As such, they predict no CP-violating effects. Should
a nonzero CP-violating observable be measured, this would rule out
these models, or at least force them to be modified.

Conclusions about the type of NP present depend on which nonzero
observables are measured:
\begin{itemize}

\item If the angular distribution is found to include the components
  $\sin 2\theta_\ell \sin 2\theta^* \sin\chi$ and $\sin^2 \theta_\ell
  \sin^2 \theta^* \sin 2\chi$ (the top two entries in Table
  \ref{Tbl:TP}), this requires a nonzero $g_R$. This can only arise in
  a $W'$ model, and so excludes all LQ models. And note: this even
  excludes the standard $W'$ models, with only a $W'_L$ or a $W'_R$. In
  this case, an unusual model, including both $W'_L$ and $W'_R$, is
  required.

\item If the $\sin 2\theta_\ell \sin 2\theta^* \sin\chi$ and $\sin^2
  \theta_\ell \sin^2 \theta^* \sin 2\chi$ components do not appear in
  the angular distribution, but $\sin\theta_\ell \sin
  2\theta^*\sin\chi$ (the third entry in Table \ref{Tbl:TP}) does,
  this indicates that $g_P$ and $g_T$ are nonzero, and that they have
  a relative phase. This can only occur in a model with two LQs. $g_T$
  can come from a $R_2$ or $S_1$ LQ, while $g_P$ can be due to a
  $U_1$, $R_2$, $S_1$ or $V_2$ LQ (but the two LQs must be different).

\item If none of the above three angular functions are present in the
  angular distribution, this implies that $g_R$ and one of $g_P$ and
  $g_T$ are zero (or that there is no phase difference). There can
  still be a CP-violating observable in the data suppressed by
  $m_{\ell}/\sqrt{q^2}$ (entries 5-8 in Table \ref{Tbl:TP}). If this
  is found to be nonzero, it does, this indicates that one of $g_T$ or
  $g_P$ (or both, if they have the same phase) is nonzero. The $g_P$
  option is particularly interesting. The $U_1$ LQ is a very popular
  NP choice (for example, see Ref.~\cite{Kumar:2018kmr}), and it can
  generate $g_P$, but not $g_T$. If this is the only nonzero
  CP-violating observable found, this would be strong support for the
  $U_1$ LQ.

\item There is also information from the CP-conserving observables.
  The full angular distribution has components proportional to
  $|\mAparT|^2$, $|\mAperpT|^2$, $|\mA0T|^2$ and $|\cA_{SP}|^2$.
  Measurements of these quantities also gives information about which
  of $g_T$ and/or $g_P$ is or is not nonzero.

\end{itemize}

\section{Conclusions}

At the present time, the anomalies in the measurements of
$R_{D^{(*)}}$ and $R_{J/\psi}$ suggest the presence of new physics in
$\bctaunu$ decays. A number of different NP explanations have been
proposed, as well as several methods for differentiating these NP
models. In this paper, we explore the possibility of using
CP-violating observables to distinguish the various NP scenarios.

The angular distribution in ${\bar B}^0 \to D^{*+} (\to D^0 \pi^+)
\tau^- {\bar\nu}_\tau$ can be used to provide CP-violating
asymmetries.  Now, the reconstruction of this angular distribution
requires the knowledge of the 3-momentum of the $\tau$. The problem
here is that ${\vec p}_\tau$ cannot be measured since its decay
products include $\nu_\tau$, which is undetected.  Thus, while our
ultimate goal is to compute the complete angular distribution,
including information related to the decay products of the $\tau$, in
this paper we take a first step by focusing on the decay
$\BDstarmunumu$. Here ${\vec p}_\mu$ is measurable, so the angular
distribution can be constructed. In addition, NP that contributes to
$\bctaunu$ may well also affect $\bcmunu$. Finally, LHCb has announced
that it intends to measure the CP-violating angular asymmetries in
$\BDstarmunumu$, and we want to examine what the implications of these
measurements are for NP.

In the SM, the hadronic $b \to c$ current is purely LH. In the
presence of NP, there can be additional contributions to this LH
current, parametrized by $g_L$, as well as other Lorentz structures:
RH ($g_R$), scalar ($g_S$), pseudoscalar ($g_P$) and tensor ($g_T$)
currents. We compute the angular distribution of $\BDstarlnu$ in terms
of the helicity amplitudes $A_i$, both in the SM and with NP. We
identify the CP-violating angular asymmetries, proportional to ${\rm
  Im}[A_i A_j^*]$, and show how all CP-violating observables depend on
any pair of $\{ (1 + g_L), g_R, g_P, g_T \}$.

We then examine the models that contribute to $\bcmunumu$. There are
two classes, involving (i) a $W'$ (two types) or (ii) a LQ (six
types). While most models contribute to $g_L$, $g_R$ can only arise in
$W'$ models, and $g_P$ and $g_T$ can only be generated due to LQ
exchange. Furthermore, not all $W'$ models lead to a nonzero $g_R$,
and not all LQ models produce $g_P$ and/or $g_T$.

The most popular explanations of the $B$ anomalies involve NP that
contributes only to $g_L$. Should any nonzero CP-violating observable
be measured, this would rule out these models, or at least require
them to be modified. In addition, there are CP-violating asymmetries
that depend on $(1 + g_L)$-$g_R$, $g_P$-$g_T$, $(1 + g_L + g_R)$-$g_P$
and $(1 + g_L - g_R)$-$g_T$ interference. By measuring all of these,
along with the CP-conserving components of the angular distribution,
it will be possible to distinguish the $W'$ and LQ models, and to
differentiate among several LQ models.

\bigskip
{\bf Acknowledgments}: We thank N. Boisvert Beaudry, K. Leblanc and
R. Watanabe for collaboration in the early stages of this project.
We thank B. Dey for a conversation regarding the experimental implementation
of the angular distribution. This work was financially supported in part
by Lawrence Technological University's Faculty Seed Grant (BB,SK) and by
NSERC of Canada (DL). BB and AD are grateful to the Mainz Institute for
Theoretical Physics (MITP) for its hospitality and its partial support
during the completion of this work. SK acknowledges the hospitality of
Lawrence Technological University, where part of this work was done.

\appendix

\section{\boldmath $\lp\cM^{\SM + \NP}\rp^2$: leptonic contributions}
\label{AppendixA}

\begin{enumerate}

\item $|\cM_{SP}|^2$:
\bea
\sum_{spins}\cL_{SP} \, \cL_{SP}^{*} &=& {\rm Tr}[(\slashed{p}_\ell+m_\ell)P_L\slashed{p}_
{\bar{\nu}}P_R] ~,
\eea
where $q = p_\ell + p_{{\bar\nu}_\ell}$.

\item $\lp\cM_{VA}\rp^2$:
\bea
\sum\limits_{\rm spins}\cL_{VA}(n)\cL^*_{VA}(n') &=& \ep^{\mu}_{VA}(n) \, \ep^{*\nu}_{VA}(n')
\, {\Tr}\[{\bar u}_{\ell}\ga_\mu P_L v_{{\bar\nu}_\ell}{\bar v}_{{\bar\nu}_\ell}\ga_\nu P_L u_{\ell}\] ~.
\label{MVAsquaredlep}
\eea

\item $\lp\cM_T\rp^2$:
\bea
\sum\limits_{\rm spins} \cL_{T}(n,p) \, \cL_{T}^*(n',p')^* &=& {\Tr}\[(\slashed{p}_\ell
+ m_\ell) \sigma_{\mu\nu} P_L \slashed{p}_{{\bar\nu}_\ell}\sigma_{\alpha\beta}P_R\] ~~~ \nl
&& \hskip1truecm \times~\ep_{T}^{\mu}(n) \, \ep_{T}^{\nu}(p) \, \ep_{T}^{*\al}(n') \, \ep_{T}^{*\be}
(p') ~.
\eea

\item $\cM_{SP}\cM^*_{VA}$:
\bea
\sum_{\rm spins}\cL_{SP} \, \cL_{VA}^*(n) &=& {\Tr}[(\slashed{p}_\ell + m_\ell) P_L \slashed{p}
_{{\bar\nu}_\ell}\gamma_\mu P_L]\epsilon_{VA}^{*\mu}(n) ~.
\label{MSPVAlep}
\eea

\item $\cM_{SP}\cM^*_T$:
\bea
\sum_{\rm spins} \cL_{SP} \, \cL_T^*(n,p) &=& i{\Tr}\[(\slashed{p}_\ell + m_\ell) P_L
\slashed{p}_{{\bar\nu}_\ell}\si_{\mu\nu}P_R\] \ep_{T}^{*\mu}(n) \, \epsilon_{T}^{*\nu}(p) ~.
\label{MSPTlep}
\eea

\item $\cM_{VA} \, \cM^*_T$:
\bea
\sum_{\rm spins}\cL_{VA}(n) \, \cL_{T}^*(n',p') &=& i{\Tr}\[(\slashed{p}_\ell + m_\ell)\ga_\mu
P_L \slashed{p}_{{\bar\nu}_\ell}\si_{\al\be}P_R\] \nl
&& \hskip1truecm \times~ \ep_{VA}^{\mu}(n)\ep_{T}^{*\al}(n') \ep_{T}^{*\be}(p') ~.
\label{MVATlep}
\eea

\end{enumerate}

The CP-violating angular asymmetries that appear in Tables \ref{N1},
\ref{N2} and \ref{N3} have two things in common: they are all
proportional to $\sin\chi$, and their coefficients are of the form
${\rm Im}({\mathcal{A}}_i {\mathcal{A}}^*_j)$, $i\ne j$. These can be
understood from the above traces. First, in the momenta, the only
element that contains $\sin\chi$ is the $y$-component of $p_\ell$
[Eq.~(\ref{pelldef})]. Second, in the evaluation of the traces, some
terms contain a factor $i$, so that ${\rm Re}({\mathcal{A}}_i i
{\mathcal{A}}^*_j) \propto{\rm Im}({\mathcal{A}}_i
{\mathcal{A}}^*_j)$. These terms come in two types. (i) In the $\perp$
polarizations, the $y$-component includes an $i$ [e.g., see
  Eq.~(\ref{poldefs})], so that $p_\ell \cdot \epsilon_{N^*}(n)$
contains $i\sin\chi$. (ii) Traces involving $\gamma_5$ lead to terms
of the form $i \epsilon_{\mu\nu\rho\si}p^\mu_\ell V_1^\nu V_2^\rho
V_3^\si$, where the $V_i$ are all different and are $\in \{q,
\epsilon_{N^*}(n) \}$ (these lead to triple-product asymmetries). If
$\mu = 2$, the factor $i\sin\chi$ is generated.

Eq.~(\ref{MSPVAlep}) contains a term of type (i) (with $N = VA$), and
leads to ${\rm Im}(\mAperp \cA_{SP}^*)$.  Eq.~(\ref{MSPTlep}) contains
a term of type (ii) (with $V_1 = q$, $V_2 = \epsilon^*_{T}(n)$, $V_3 =
\epsilon^*_{T}(n')$), and leads to ${\rm Im}(\cA_{SP} \mAperpT^*)$.
Eq.~(\ref{MVAsquaredlep}) contains both type (i) (with $N = VA$),
leading to ${\rm Im}(\mAn \mApar^*)$, and type (ii) (with $V_1 = q$,
$V_2 = \epsilon_{VA}(n)$, $V_3 = \epsilon^*_{VA}(n')$), leading to
${\rm Im}(\mAperp \mAn^*)$, ${\rm Im}(\mApar \mAperp^*)$ and ${\rm
  Im}(\mAt \mAperp^*)$.  Eq.~(\ref{MVATlep}) contains both type (i)
(with $N = T$), leading to ${\rm Im}(\mAn \mAparT^*)$ and ${\rm
  Im}(\mApar \mA0T^*)$, and type (ii) (with $V_1 = \epsilon_{VA}(m)$,
$V_2 = \epsilon^*_{T}(n)$, $V_3 = \epsilon^*_{T}(n')$), leading to
${\rm Im}(\mAt \mAperpT^*)$.

\section{Helicity amplitudes in terms of form factors}
\label{AppendixB}

Using the definitions for the $B \to D^*$ form factors given in
Refs.~\cite{Sakaki:2013bfa, Beneke:2000wa}, we can find the hadronic
helicity amplitudes [Eq.~(\ref{eq:Tampdef})]:
\begin{align}
\mathcal{A}_{SP} &= -g_P \, \frac{\sqrt{\lambda(m_B^2,\mDs^2, q^2)}}{m_b+m_c} A_0(q^2) ~, \nonumber \\[5pt]
\mathcal{A}_{0} &= -(1+g_L-g_R) \, \frac{(m_B + \mDs)(m_B^2-\mDs^2 -q^2)}{2\mDs \sqrt{q^2}} A_1(q^2) \nl
& \hskip1truecm
~+ (1+g_L-g_R) \, \frac{\lambda(m_B^2, \mDs^2 , q^2)}{2\mDs (m_B + \mDs) \sqrt{q^2}} A_2(q^2) ~, \nonumber \\[5pt]
\mathcal{A}_{t} &= -(1+g_L-g_R) \, \frac{\sqrt{\lambda(m_B^2,\mDs^2, q^2)}}{\sqrt{q^2}} A_0(q^2) ~, \nonumber \\[5pt]
\mathcal{A}_{+} &= (1+g_L-g_R) \, (m_B + \mDs)A_1(q^2) - (1+g_L+g_R)\frac{\sqrt{\lambda(m_B^2,\mDs^2, q^2)}}{m_B + \mDs} V(q^2)
~, \nonumber \\[5pt]
\mathcal{A}_{-} &= (1+g_L-g_R) \, (m_B + \mDs)A_1(q^2) + (1+g_L+g_R)\frac{\sqrt{\lambda(m_B^2,\mDs^2, q^2)}}{m_B + \mDs}V(q^2) ~,
\nonumber \\[5pt]
\mA0T &= g_{T} \, \frac{1}{2\mDs(m_B^2-\mDs^2)} \Big( (m_B^2-\mDs^2)(m_B^2+3\mDs^2 -q^2)T_2(q^2) -\lambda(m_B^2,\mDs^2, q^2)T_3(q^2) \Big)
~, \nonumber \\[5pt]
\mathcal{A}_{\pm, T} &=  g_{T} \, \frac{\sqrt{\lambda(m_B^2,\mDs^2, q^2)}T_1(q^2) \pm (m_B^2-\mDs^2)T_2(q^2)}{\sqrt{q^2}} ~,
\end{align}
where $\lambda(a,b,c)=a^2+b^2+c^2-2ab-2ac-2bc$.


\begin{thebibliography}{99}

\bibitem{RD_BaBar}
  J.~P.~Lees {\it et al.} [BaBar Collaboration],
  ``Measurement of an Excess of $\bar{B} \to D^{(*)}\tau^- \bar{\nu}_\tau$ Decays and Implications for Charged Higgs Bosons,''
  Phys.\ Rev.\ D {\bf 88}, 072012 (2013)
  doi:10.1103/PhysRevD.88.072012
  [arXiv:1303.0571 [hep-ex]].

\bibitem{RD_Belle}
M.~Huschle {\it et al.} [Belle Collaboration],
  ``Measurement of the branching ratio of $\bar{B} \to D^{(\ast)} \tau^- \bar{\nu}_\tau$ relative to $\bar{B} \to D^{(\ast)} \ell^- \bar{\nu}_\ell$ decays with hadronic tagging at Belle,''
  Phys.\ Rev.\ D {\bf 92}, 072014 (2015)
  doi:10.1103/PhysRevD.92.072014
  [arXiv:1507.03233 [hep-ex]].

\bibitem{RD_LHCb}
R.~Aaij {\it et al.} [LHCb Collaboration],
  ``Measurement of the ratio of branching fractions $\mathcal{B}(\bar{B}^0 \to D^{*+}\tau^{-}\bar{\nu}_{\tau})/\mathcal{B}(\bar{B}^0 \to D^{*+}\mu^{-}\bar{\nu}_{\mu})$,''
  Phys.\ Rev.\ Lett.\  {\bf 115}, 111803 (2015)
  Addendum: [Phys.\ Rev.\ Lett.\  {\bf 115}, 159901 (2015)]
  doi:10.1103/PhysRevLett.115.159901, 10.1103/PhysRevLett.115.111803
  [arXiv:1506.08614 [hep-ex]].

\bibitem{Abdesselam:2016xqt}
  A.~Abdesselam {\it et al.},
  ``Measurement of the $\tau$ lepton polarization in the decay ${\bar B} \rightarrow D^* \tau^- {\bar \nu_{\tau}}$,''
  arXiv:1608.06391 [hep-ex].

\bibitem{Aaij:2017tyk}
  R.~Aaij {\it et al.} [LHCb Collaboration],
  ``Measurement of the ratio of branching fractions $\mathcal{B}(B_c^+\,\to\,J/\psi\tau^+\nu_\tau)$/$\mathcal{B}(B_c^+\,\to\,J/\psi\mu^+\nu_\mu)$,''
  Phys.\ Rev.\ Lett.\  {\bf 120}, no. 12, 121801 (2018)
  doi:10.1103/PhysRevLett.120.121801
  [arXiv:1711.05623 [hep-ex]].

\bibitem{Abdesselam:2017kjf}
  A.~Abdesselam {\it et al.} [Belle Collaboration],
  ``Precise determination of the CKM matrix element $\left| V_{cb}\right|$ with $\bar B^0 \to D^{*\,+} \, \ell^- \, \bar \nu_\ell$ decays with hadronic tagging at Belle,''
  arXiv:1702.01521 [hep-ex].

\bibitem{Bernlochner:2017jka}
  F.~U.~Bernlochner, Z.~Ligeti, M.~Papucci and D.~J.~Robinson,
  ``Combined analysis of semileptonic $B$ decays to $D$ and $D^*$: $R(D^{(*)})$, $|V_{cb}|$, and new physics,''
  Phys.\ Rev.\ D {\bf 95}, no. 11, 115008 (2017)
  Erratum: [Phys.\ Rev.\ D {\bf 97}, no. 5, 059902 (2018)]
  doi:10.1103/PhysRevD.95.115008, 10.1103/PhysRevD.97.059902
  [arXiv:1703.05330 [hep-ph]].

\bibitem{Bigi:2017jbd}
  D.~Bigi, P.~Gambino and S.~Schacht,
  ``$R(D^*)$, $|V_{cb}|$, and the Heavy Quark Symmetry relations between form factors,''
  JHEP {\bf 1711}, 061 (2017)
  doi:10.1007/JHEP11(2017)061
  [arXiv:1707.09509 [hep-ph]].

\bibitem{Jaiswal:2017rve}
  S.~Jaiswal, S.~Nandi and S.~K.~Patra,
  ``Extraction of $|V_{cb}|$ from $B\to D^{(*)}\ell\nu_\ell$ and the Standard Model predictions of $R(D^{(*)})$,''
  JHEP {\bf 1712}, 060 (2017)
  doi:10.1007/JHEP12(2017)060
  [arXiv:1707.09977 [hep-ph]].

 \bibitem{Watanabe:2017mip}
  R.~Watanabe,
  ``New Physics effect on $B_c \to J/\psi \tau\bar\nu$ in relation to the $R_{D^{(*)}}$ anomaly,''
  Phys.\ Lett.\ B {\bf 776}, 5 (2018)
  doi:10.1016/j.physletb.2017.11.016
  [arXiv:1709.08644 [hep-ph]].

\bibitem{Fajfer:2012jt}
  S.~Fajfer, J.~F.~Kamenik, I.~Nisandzic and J.~Zupan,
  ``Implications of Lepton Flavor Universality Violations in $B$ Decays,''
  Phys.\ Rev.\ Lett.\  {\bf 109}, 161801 (2012)
  doi:10.1103/PhysRevLett.109.161801
  [arXiv:1206.1872 [hep-ph]].

\bibitem{Datta:2012qk}
  A.~Datta, M.~Duraisamy and D.~Ghosh,
  ``Diagnosing New Physics in $b \to c \, \tau \, \nu_\tau$ decays in the light of the recent BaBar result,''
  Phys.\ Rev.\ D {\bf 86}, 034027 (2012)
  doi:10.1103/PhysRevD.86.034027
  [arXiv:1206.3760 [hep-ph]].


\bibitem{Tanaka:2012nw}
  M.~Tanaka and R.~Watanabe,
  ``New physics in the weak interaction of $\bar B\to D^{(*)}\tau\bar\nu$,''
  Phys.\ Rev.\ D {\bf 87}, no. 3, 034028 (2013)
  doi:10.1103/PhysRevD.87.034028
  [arXiv:1212.1878 [hep-ph]].

\bibitem{Biancofiore:2013ki}
  P.~Biancofiore, P.~Colangelo and F.~De Fazio,
  ``On the anomalous enhancement observed in $B \to D^{(*)}\tau{\bar \nu}_\tau$ decays,''
  Phys.\ Rev.\ D {\bf 87}, no. 7, 074010 (2013)
  doi:10.1103/PhysRevD.87.074010
  [arXiv:1302.1042 [hep-ph]].

  \bibitem{Duraisamy:2013kcw}
  M.~Duraisamy and A.~Datta,
  ``The Full $B \to D^{*} \tau^{-} \bar{\nu_\tau}$ Angular Distribution and CP violating Triple Products,''
  JHEP {\bf 1309}, 059 (2013)
  doi:10.1007/JHEP09(2013)059
  [arXiv:1302.7031 [hep-ph]].




\bibitem{Freytsis:2015qca}
  M.~Freytsis, Z.~Ligeti and J.~T.~Ruderman,
  ``Flavor models for $\bar{B} \to D^{(*)} \tau \bar{\nu}$,''
  Phys.\ Rev.\ D {\bf 92}, no. 5, 054018 (2015)
  doi:10.1103/PhysRevD.92.054018
  [arXiv:1506.08896 [hep-ph]].

\bibitem{Bardhan:2016uhr}
  D.~Bardhan, P.~Byakti and D.~Ghosh,
  ``A closer look at the $R_{D}$ and $R_{D^*}$ anomalies,''
  JHEP {\bf 1701}, 125 (2017)
  doi:10.1007/JHEP01(2017)125
  [arXiv:1610.03038 [hep-ph]].

\bibitem{Bhattacharya:2016zcw}
  S.~Bhattacharya, S.~Nandi and S.~K.~Patra,
  ``Looking for possible new physics in $B\to D^{(\ast)}\tau\nu_{\tau}$ in light of recent data,''
  Phys.\ Rev.\ D {\bf 95}, no. 7, 075012 (2017)
  doi:10.1103/PhysRevD.95.075012
  [arXiv:1611.04605 [hep-ph]].

\bibitem{Dutta:2017wpq}
  R.~Dutta,
  ``Exploring $R_D$, $R_{D^{\ast}}$ and $R_{J/\Psi}$ anomalies,''
  arXiv:1710.00351 [hep-ph].

\bibitem{Alok:2017qsi}
  A.~K.~Alok, D.~Kumar, J.~Kumar, S.~Kumbhakar and S.~U.~Sankar,
  ``New physics solutions for $R_D$ and $R_{D^*}$,''
  JHEP {\bf 1809}, 152 (2018)
  doi:10.1007/JHEP09(2018)152
  [arXiv:1710.04127 [hep-ph]].

\bibitem{Huang:2018nnq}
  Z.~R.~Huang, Y.~Li, C.~D.~Lu, M.~A.~Paracha and C.~Wang,
  ``Footprints of New Physics in $b\to c\tau\nu$ Transitions,''
  Phys.\ Rev.\ D {\bf 98}, no. 9, 095018 (2018)
  doi:10.1103/PhysRevD.98.095018
  [arXiv:1808.03565 [hep-ph]].


\bibitem{Crivellin:2012ye}
  A.~Crivellin, C.~Greub and A.~Kokulu,
  ``Explaining $B\to D\tau\nu$, $B\to D^*\tau\nu$ and $B\to \tau\nu$ in a 2HDM of type III,''
  Phys.\ Rev.\ D {\bf 86}, 054014 (2012)
  doi:10.1103/PhysRevD.86.054014
  [arXiv:1206.2634 [hep-ph]].

\bibitem{Celis:2012dk}
  A.~Celis, M.~Jung, X.~Q.~Li and A.~Pich,
  ``Sensitivity to charged scalars in $\boldmath{B\to D^{(*)}\tau\nu_\tau}$ and $\boldmath{B\to\tau\nu_\tau}$ decays,''
  JHEP {\bf 1301}, 054 (2013)
  doi:10.1007/JHEP01(2013)054
  [arXiv:1210.8443 [hep-ph]].

\bibitem{He:2012zp}
  X.~G.~He and G.~Valenia,
  ``$B$ decays with $\tau$ leptons in nonuniversal left-right models,''
  Phys.\ Rev.\ D {\bf 87}, no. 1, 014014 (2013)
  doi:10.1103/PhysRevD.87.014014
  [arXiv:1211.0348 [hep-ph]].

\bibitem{Ko:2012sv}
  P.~Ko, Y.~Omura and C.~Yu,
  ``$B \to D^{(*)}$ tau nu and $B \to$ tau nu in chiral $U(1)'$ models with flavored multi Higgs doublets,''
  JHEP {\bf 1303}, 151 (2013)
  doi:10.1007/JHEP03(2013)151
  [arXiv:1212.4607 [hep-ph]].

\bibitem{Dorsner:2013tla}
  I.~Doršner, S.~Fajfer, N.~Košnik and I.~Nišandžić,
  ``Minimally flavored colored scalar in $\bar B \to D^{(*)} \tau \bar \nu$ and the mass matrices constraints,''
  JHEP {\bf 1311}, 084 (2013)
  doi:10.1007/JHEP11(2013)084
  [arXiv:1306.6493 [hep-ph]].

\bibitem{Sakaki:2013bfa}
  Y.~Sakaki, M.~Tanaka, A.~Tayduganov and R.~Watanabe,
  ``Testing leptoquark models in $\bar B \to D^{(*)} \tau \bar\nu$,''
  Phys.\ Rev.\ D {\bf 88}, no. 9, 094012 (2013)
  doi:10.1103/PhysRevD.88.094012
  [arXiv:1309.0301 [hep-ph]].

\bibitem{Greljo:2015mma}
  A.~Greljo, G.~Isidori and D.~Marzocca,
  ``On the breaking of Lepton Flavor Universality in $B$ decays,''
  JHEP {\bf 1507}, 142 (2015)
  doi:10.1007/JHEP07(2015)142
  [arXiv:1506.01705 [hep-ph]].

\bibitem{Crivellin:2015hha}
  A.~Crivellin, J.~Heeck and P.~Stoffer,
  ``A perturbed lepton-specific two-Higgs-doublet model facing experimental hints for physics beyond the Standard Model,''
  Phys.\ Rev.\ Lett.\  {\bf 116}, no. 8, 081801 (2016)
  doi:10.1103/PhysRevLett.116.081801
  [arXiv:1507.07567 [hep-ph]].

\bibitem{Dumont:2016xpj}
  B.~Dumont, K.~Nishiwaki and R.~Watanabe,
  ``LHC constraints and prospects for $S_1$ scalar leptoquark explaining the $\bar B \to D^{(*)} \tau \bar\nu$ anomaly,''
  Phys.\ Rev.\ D {\bf 94}, no. 3, 034001 (2016)
  doi:10.1103/PhysRevD.94.034001
  [arXiv:1603.05248 [hep-ph]].

\bibitem{Boucenna:2016wpr}
  S.~M.~Boucenna, A.~Celis, J.~Fuentes-Martin, A.~Vicente and J.~Virto,
  ``Non-abelian gauge extensions for $B$-decay anomalies,''
  Phys.\ Lett.\ B {\bf 760}, 214 (2016)
  doi:10.1016/j.physletb.2016.06.067
  [arXiv:1604.03088 [hep-ph]].

\bibitem{Boucenna:2016qad}
  S.~M.~Boucenna, A.~Celis, J.~Fuentes-Martin, A.~Vicente and J.~Virto,
  ``Phenomenology of an $SU(2) \times SU(2) \times U(1)$ model with lepton-flavour non-universality,''
  JHEP {\bf 1612}, 059 (2016)
  doi:10.1007/JHEP12(2016)059
  [arXiv:1608.01349 [hep-ph]].

\bibitem{Bhattacharya:2016mcc}
  B.~Bhattacharya, A.~Datta, J.~P.~Gu\'evin, D.~London and R.~Watanabe,
  ``Simultaneous Explanation of the $R_K$ and $R_{D^{(*)}}$ Puzzles: a Model Analysis,''
  JHEP {\bf 1701}, 015 (2017)
  doi:10.1007/JHEP01(2017)015
  [arXiv:1609.09078 [hep-ph]].

\bibitem{Alonso:2016oyd}
  R.~Alonso, B.~Grinstein and J.~Martin Camalich,
  ``Lifetime of $B_c^-$ Constrains Explanations for Anomalies in  $B\to D^{(*)}\tau\nu$,''
  Phys.\ Rev.\ Lett.\  {\bf 118}, no. 8, 081802 (2017)
  doi:10.1103/PhysRevLett.118.081802
  [arXiv:1611.06676 [hep-ph]].

\bibitem{Celis:2016azn}
  A.~Celis, M.~Jung, X.~Q.~Li and A.~Pich,
  ``Scalar contributions to $b\to c (u) \tau \nu$ transitions,''
  Phys.\ Lett.\ B {\bf 771}, 168 (2017)
  doi:10.1016/j.physletb.2017.05.037
  [arXiv:1612.07757 [hep-ph]].

\bibitem{Wei:2017ago}
  M.~Wei and Y.~Chong-Xing,
  ``Charged Higgs bosons from the 3-3-1 models and the $\mathcal{R}(D^{(*)})$ anomalies,''
  Phys.\ Rev.\ D {\bf 95}, no. 3, 035040 (2017)
  doi:10.1103/PhysRevD.95.035040
  [arXiv:1702.01255 [hep-ph]].

\bibitem{Altmannshofer:2017poe}
  W.~Altmannshofer, P.~S.~Bhupal Dev and A.~Soni,
  ``$R_{D^{(*)}}$ anomaly: A possible hint for natural supersymmetry with $R$-parity violation,''
  Phys.\ Rev.\ D {\bf 96}, no. 9, 095010 (2017)
  doi:10.1103/PhysRevD.96.095010
  [arXiv:1704.06659 [hep-ph]].

\bibitem{Buttazzo:2017ixm}
  D.~Buttazzo, A.~Greljo, G.~Isidori and D.~Marzocca,
  ``$B$-physics anomalies: a guide to combined explanations,''
  JHEP {\bf 1711}, 044 (2017)
  doi:10.1007/JHEP11(2017)044
  [arXiv:1706.07808 [hep-ph]].

\bibitem{Iguro:2017ysu}
  S.~Iguro and K.~Tobe,
  ``$R(D^{(*)})$ in a general two Higgs doublet model,''
  Nucl.\ Phys.\ B {\bf 925}, 560 (2017)
  doi:10.1016/j.nuclphysb.2017.10.014
  [arXiv:1708.06176 [hep-ph]].

\bibitem{He:2017bft}
  X.~G.~He and G.~Valencia,
  ``Lepton universality violation and right-handed currents in $b \to c \tau \nu$,''
  Phys.\ Lett.\ B {\bf 779}, 52 (2018)
  doi:10.1016/j.physletb.2018.01.073
  [arXiv:1711.09525 [hep-ph]].

\bibitem{Biswas:2018jun}
  A.~Biswas, D.~K.~Ghosh, A.~Shaw and S.~K.~Patra,
  ``$b \to c \ell \nu$ anomalies in light of extended scalar sectors,''
  arXiv:1801.03375 [hep-ph].

\bibitem{Asadi:2018wea}
  P.~Asadi, M.~R.~Buckley and D.~Shih,
  ``It's all right(-handed neutrinos): a new W$^{\prime}$ model for the $ {R}_{D^{{\left(\ast \right)}}} $ anomaly,''
  JHEP {\bf 1809}, 010 (2018)
  doi:10.1007/JHEP09(2018)010
  [arXiv:1804.04135 [hep-ph]].

\bibitem{Greljo:2018ogz}
  A.~Greljo, D.~J.~Robinson, B.~Shakya and J.~Zupan,
  ``R(D$^{(*)}$) from $W'$ and right-handed neutrinos,''
  JHEP {\bf 1809}, 169 (2018)
  doi:10.1007/JHEP09(2018)169
  [arXiv:1804.04642 [hep-ph]].

\bibitem{Azatov:2018knx}
  A.~Azatov, D.~Bardhan, D.~Ghosh, F.~Sgarlata and E.~Venturini,
  ``Anatomy of $b \to c \tau \nu$ anomalies,''
  JHEP {\bf 1811}, 187 (2018)
  doi:10.1007/JHEP11(2018)187
  [arXiv:1805.03209 [hep-ph]].

\bibitem{Martinez:2018ynq}
  R.~Martinez, C.~F.~Sierra and G.~Valencia,
  ``Beyond $\mathcal{R}(D^{(*)})$ with the general type-III 2HDM for $b\to c\tau\nu$,''
  Phys.\ Rev.\ D {\bf 98}, no. 11, 115012 (2018)
  doi:10.1103/PhysRevD.98.115012
  [arXiv:1805.04098 [hep-ph]].

\bibitem{Fraser:2018aqj}
  S.~Fraser, C.~Marzo, L.~Marzola, M.~Raidal and C.~Spethmann,
  ``Towards a viable scalar interpretation of $R_{D^{(*)}}$,''
  Phys.\ Rev.\ D {\bf 98}, no. 3, 035016 (2018)
  doi:10.1103/PhysRevD.98.035016
  [arXiv:1805.08189 [hep-ph]].

\bibitem{Kumar:2018kmr}
  J.~Kumar, D.~London and R.~Watanabe,
  ``Combined Explanations of the $b \to s \mu^+ \mu^-$ and $b \to c \tau^- {\bar\nu}$ Anomalies: a General Model Analysis,''
  Phys.\ Rev.\ D {\bf 99}, no. 1, 015007 (2019)
  doi:10.1103/PhysRevD.99.015007
  [arXiv:1806.07403 [hep-ph]].

\bibitem{Robinson:2018gza}
  D.~J.~Robinson, B.~Shakya and J.~Zupan,
  ``Right-handed Neutrinos and $R(D^{(*)})$,''
  arXiv:1807.04753 [hep-ph].

\bibitem{Marzo:2019ldg}
  C.~Marzo, L.~Marzola and M.~Raidal,
  ``Common explanation to the $R_{K^{(*)}}$, $R_{D^{(*)}}$ and $\epsilon^\prime/\epsilon$ anomalies in a 3HDM+$\nu_R$ and connections to neutrino physics,''
  arXiv:1901.08290 [hep-ph].


\bibitem{Sakaki:2012ft}
  Y.~Sakaki and H.~Tanaka,
  ``Constraints on the charged scalar effects using the forward-backward asymmetry on $B^- \to D^{(*)} \tau^- {\bar\nu}_\tau$,''
  Phys.\ Rev.\ D {\bf 87}, no. 5, 054002 (2013)
  doi:10.1103/PhysRevD.87.054002
  [arXiv:1205.4908 [hep-ph]].

\bibitem{Duraisamy:2014sna}
  M.~Duraisamy, P.~Sharma and A.~Datta,
  ``Azimuthal $B \to D^{*} \tau^{-} \bar{\nu_\tau}$ angular distribution with tensor operators,''
  Phys.\ Rev.\ D {\bf 90}, no. 7, 074013 (2014)
  doi:10.1103/PhysRevD.90.074013
  [arXiv:1405.3719 [hep-ph]].

\bibitem{Sakaki:2014sea}
  Y.~Sakaki, M.~Tanaka, A.~Tayduganov and R.~Watanabe,
  ``Probing New Physics with $q^2$ distributions in $\bar{B} \to D^{(*)} \tau \bar\nu$,''
  Phys.\ Rev.\ D {\bf 91}, no. 11, 114028 (2015)
  doi:10.1103/PhysRevD.91.114028
  [arXiv:1412.3761 [hep-ph]].

\bibitem{Bhattacharya:2015ida}
  S.~Bhattacharya, S.~Nandi and S.~K.~Patra,
  ``Optimal-observable analysis of possible new physics in $B\to D^{(\ast)}\tau\nu_{\tau}$,''
  Phys.\ Rev.\ D {\bf 93}, no. 3, 034011 (2016)
  doi:10.1103/PhysRevD.93.034011
  [arXiv:1509.07259 [hep-ph]].

\bibitem{Alonso:2016gym}
  R.~Alonso, A.~Kobach and J.~Martin Camalich,
  ``New physics in the kinematic distributions of $\bar B\to D^{(*)}\tau^-(\to\ell^-\bar\nu_\ell\nu_\tau)\bar\nu_\tau$,''
  Phys.\ Rev.\ D {\bf 94}, no. 9, 094021 (2016)
  doi:10.1103/PhysRevD.94.094021
  [arXiv:1602.07671 [hep-ph]].

\bibitem{Alok:2016qyh}
  A.~K.~Alok, D.~Kumar, S.~Kumbhakar and S.~U.~Sankar,
  ``$D^{*}$ polarization as a probe to discriminate new physics in $\bar{B}\to D^{*} \tau \bar{\nu}$,''
  Phys.\ Rev.\ D {\bf 95}, no. 11, 115038 (2017)
  doi:10.1103/PhysRevD.95.115038
  [arXiv:1606.03164 [hep-ph]].

\bibitem{Ligeti:2016npd}
  Z.~Ligeti, M.~Papucci and D.~J.~Robinson,
  ``New Physics in the Visible Final States of $B\to D^{(*)}\tau\nu$,''
  JHEP {\bf 1701}, 083 (2017)
  doi:10.1007/JHEP01(2017)083
  [arXiv:1610.02045 [hep-ph]].

\bibitem{Ivanov:2017mrj}
  M.~A.~Ivanov, J.~G.~K\"orner and C.~T.~Tran,
  ``Probing new physics in $\bar{B}^0 \to D^{(\ast)} \tau^- \bar\nu_{\tau}$ using the longitudinal, transverse, and normal polarization components of the tau lepton,''
  Phys.\ Rev.\ D {\bf 95}, no. 3, 036021 (2017)
  doi:10.1103/PhysRevD.95.036021
  [arXiv:1701.02937 [hep-ph]].

\bibitem{Aloni:2017eny}
  D.~Aloni, A.~Efrati, Y.~Grossman and Y.~Nir,
  ``$\Upsilon$ and $\psi$ leptonic decays as probes of solutions to the $R_D^{(*)}$ puzzle,''
  JHEP {\bf 1706}, 019 (2017)
  doi:10.1007/JHEP06(2017)019
  [arXiv:1702.07356 [hep-ph]].

\bibitem{Colangelo:2018cnj}
  P.~Colangelo and F.~De Fazio,
  ``Scrutinizing $ \overline{B}\to {D}^{\ast}\left(D\pi \right){\ell}^{-}{\overline{\nu}}_{\ell } $ and $ \overline{B}\to {D}^{\ast}\left(D\gamma \right){\ell}^{-}{\overline{\nu}}_{\ell } $ in search of new physics footprints,''
  JHEP {\bf 1806}, 082 (2018)
  doi:10.1007/JHEP06(2018)082
  [arXiv:1801.10468 [hep-ph]].

\bibitem{Alok:2018uft}
  A.~K.~Alok, D.~Kumar, S.~Kumbhakar and S.~Uma Sankar,
  ``Resolution of $R_D$/$R_{D^*}$ puzzle,''
  Phys.\ Lett.\ B {\bf 784}, 16 (2018)
  doi:10.1016/j.physletb.2018.07.001
  [arXiv:1804.08078 [hep-ph]].

\bibitem{Aloni:2018ipm}
  D.~Aloni, Y.~Grossman and A.~Soffer,
  ``Measuring CP violation in $b\to c\tau^-\bar{\nu}_\tau$ using excited charm mesons,''
  Phys.\ Rev.\ D {\bf 98}, no. 3, 035022 (2018)
  doi:10.1103/PhysRevD.98.035022
  [arXiv:1806.04146 [hep-ph]].

\bibitem{Asadi:2018sym}
  P.~Asadi, M.~R.~Buckley and D.~Shih,
  ``Asymmetry Observables and the Origin of $R_{D^{(*)}}$ Anomalies,''
  arXiv:1810.06597 [hep-ph].

\bibitem{Blanke:2018yud}
  M.~Blanke, A.~Crivellin, S.~de Boer, M.~Moscati, U.~Nierste, I.~Nišandžić and T.~Kitahara,
  ``Impact of polarization observables and $ B_c\to \tau \nu$ on new physics explanations of the $b\to c \tau \nu$ anomaly,''
  arXiv:1811.09603 [hep-ph].

\bibitem{Iguro:2018vqb}
  S.~Iguro, T.~Kitahara, R.~Watanabe and K.~Yamamoto,
  ``$D^{\ast}$ polarization vs. $R_{D^{(\ast)}}$ anomalies in the leptoquark models,''
  arXiv:1811.08899 [hep-ph].

\bibitem{Datta:2004re}
  A.~Datta and D.~London,
  ``Measuring new physics parameters in $B$ penguin decays,''
  Phys.\ Lett.\ B {\bf 595}, 453 (2004)
  doi:10.1016/j.physletb.2004.06.069
  [hep-ph/0404130].

\bibitem{Alok:2017jgr}
  A.~K.~Alok, B.~Bhattacharya, D.~Kumar, J.~Kumar, D.~London and S.~U.~Sankar,
  ``New physics in $b \rightarrow s \mu^+ \mu^-$: Distinguishing models through CP-violating effects,''
  Phys.\ Rev.\ D {\bf 96}, no. 1, 015034 (2017)
  doi:10.1103/PhysRevD.96.015034
  [arXiv:1703.09247 [hep-ph]].

\bibitem{Hagiwara:2014tsa}
  K.~Hagiwara, M.~M.~Nojiri and Y.~Sakaki,
  ``$CP$ violation in $B \to D\tau \nu_{\tau}$ using multipion tau decays,''
  Phys.\ Rev.\ D {\bf 89}, no. 9, 094009 (2014)
  doi:10.1103/PhysRevD.89.094009
  [arXiv:1403.5892 [hep-ph]].

  \bibitem{Alonso:2017ktd}
  R.~Alonso, J.~Martin Camalich and S.~Westhoff,
  ``Tau properties in $B\to D\tau\nu$ from visible final-state kinematics,''
  Phys.\ Rev.\ D {\bf 95}, no. 9, 093006 (2017)
  doi:10.1103/PhysRevD.95.093006
  [arXiv:1702.02773 [hep-ph]].

\bibitem{Marangotto:2018pbs}
  D.~Marangotto,
  ``Angular and CP-violation analyses of $\bar{B}\to D^{*+} l^-\bar{\nu}_{l}$ decays at hadron collider experiments,''
  arXiv:1812.08144 [hep-ex].

\bibitem{Altmannshofer:2008dz}
  W.~Altmannshofer, P.~Ball, A.~Bharucha, A.~J.~Buras, D.~M.~Straub and M.~Wick,
  ``Symmetries and Asymmetries of $B \to K^{*} \mu^{+} \mu^{-}$ Decays in the Standard Model and Beyond,''
  JHEP {\bf 0901}, 019 (2009)
  doi:10.1088/1126-6708/2009/01/019
  [arXiv:0811.1214 [hep-ph]].

\bibitem{Dey:2015rqa}
  B.~Dey,
  ``Angular analyses of exclusive $\overline{B} \to X \ell_1 \ell_2$ with complex helicity amplitudes,''
  Phys.\ Rev.\ D {\bf 92}, 033013 (2015)
  doi:10.1103/PhysRevD.92.033013
  [arXiv:1505.02873 [hep-ex]].

\bibitem{Bobeth:2012vn}
  C.~Bobeth, G.~Hiller and D.~van Dyk,
  ``General analysis of $\bar{B} \to \bar{K}^{(*)}\ell^+ \ell^-$  decays at low recoil,''
  Phys.\ Rev.\ D {\bf 87}, no. 3, 034016 (2013)
  [Phys.\ Rev.\ D {\bf 87}, 034016 (2013)]
  doi:10.1103/PhysRevD.87.034016
  [arXiv:1212.2321 [hep-ph]].

\bibitem{TPs}
See, for example,
A.~Datta and D.~London,
  ``Triple-product correlations in $B \to V_1 V_2$ decays and new physics,''
  Int.\ J.\ Mod.\ Phys.\ A {\bf 19}, 2505 (2004)
  doi:10.1142/S0217751X04018300
  [hep-ph/0303159].

\bibitem{Gronau:2011cf}
  M.~Gronau and J.~L.~Rosner,
  ``Triple product asymmetries in $K$, $D_{(s)}$ and $B_{(s)}$ decays,''
  Phys.\ Rev.\ D {\bf 84}, 096013 (2011)
  doi:10.1103/PhysRevD.84.096013
  [arXiv:1107.1232 [hep-ph]].

\bibitem{Crivellin:2017dsk}
  A.~Crivellin, D.~Müller, A.~Signer and Y.~Ulrich,
  ``Correlating lepton flavor universality violation in $B$ decays with $\mu\to e\gamma$ using leptoquarks,''
  Phys.\ Rev.\ D {\bf 97}, no. 1, 015019 (2018)
  doi:10.1103/PhysRevD.97.015019
  [arXiv:1706.08511 [hep-ph]].

\bibitem{pdg}
M.~Tanabashi {\it et al.} [Particle Data Group],
  ``Review of Particle Physics,''
  Phys.\ Rev.\ D {\bf 98}, no. 3, 030001 (2018).
  doi:10.1103/PhysRevD.98.030001

\bibitem{RKRDpaper}
  B.~Bhattacharya, A.~Datta, D.~London and S.~Shivashankara,
  ``Simultaneous Explanation of the $R_K$ and $R(D^{(*)})$ Puzzles,''
  Phys.\ Lett.\ B {\bf 742}, 370 (2015)
  doi:10.1016/j.physletb.2015.02.011
  [arXiv:1412.7164 [hep-ph]].

\bibitem{Alok:2017sui}
  A.~K.~Alok, B.~Bhattacharya, A.~Datta, D.~Kumar, J.~Kumar and D.~London,
  ``New Physics in $b \to s \mu^+ \mu^-$ after the Measurement of $R_{K^*}$,''
  Phys.\ Rev.\ D {\bf 96}, no. 9, 095009 (2017)
  doi:10.1103/PhysRevD.96.095009
  [arXiv:1704.07397 [hep-ph]].

\bibitem{Beneke:2000wa}
  M.~Beneke and T.~Feldmann,
  ``Symmetry breaking corrections to heavy to light B meson form-factors at large recoil,''
  Nucl.\ Phys.\ B {\bf 592}, 3 (2001)
  doi:10.1016/S0550-3213(00)00585-X
  [hep-ph/0008255].

\end{thebibliography}
\end{document}